\documentclass[%
 reprint,
 amsmath,amssymb,
 aps,
]{revtex4-1}

\usepackage{graphicx}
\usepackage{dcolumn}
\usepackage{bm}
\usepackage[mathlines]{lineno}

\usepackage{color}
\usepackage{eso-pic} 

\begin{document}

\preprint{APS/123-QED}

\title{Measurement of $\nu_{\mu}$ and $\bar{\nu}_{\mu}$ Neutral Current $\pi^{0} \rightarrow \gamma\gamma$ Production in the ArgoNeuT Detector}

\author{R. Acciarri,$^{1}$ C. Adams,$^{2}$ J. Asaadi,$^{3}$ B. Baller,${1}$, T. Bolton,$^{4}$ C. Bromberg,$^{5}$ F. Cavanna,$^{1,6}$ E. Church,$^{7}$ D.~Edmunds,$^{5}$ A. Ereditato,$^{8}$ S. Farooq,$^{4}$ B. Fleming,$^{2}$ H. Greenlee,$^{1}$ A. Hackenburg$^{2}$ R.~Hatcher,$^{1}$ G.~Horton-Smith,$^{4}$ C.~James,$^{1}$ E.~Klein,$^{2}$ K. Lang,$^{9}$ P. Laurens,$^{5}$ R. Mehdiyev,$^{9}$ B. Page,$^{5}$ O. Palamara,$^{1,10}$ K. Partyka,$^{2}$ G.~Rameika,$^{1}$ B.~Rebel,$^{1}$ A. Schukraft,$^{1}$ M. Soderberg,$^{1,3}$ J. Spitz,$^{2}$ A.M Szelc,$^{2}$ M. Weber$^{8}$ T. Yang,$^{1}$ G.P. Zeller$^{1}$ \\}

\collaboration{ArgoNeuT Collaboration}

\affiliation{                                                                 
\centerline{$^{1}$Fermi National Accelerator Laboratory, Batavia, IL 60510}
\centerline{$^{2}$Yale University, New Haven, CT 06520} 
\centerline{$^{3}$Syracuse University, Syracuse, NY 13244}                                     
\centerline{$^{4}$Kansas State University, Manhattan, KS 66506}                     
\centerline{$^{5}$Michigan State University, East Lansing, MI 48824}                           
\centerline{$^{6}$Universita dell'Aquila e INFN, L'Aquila, Italy}  
\centerline{$^{7}$Pacific Northwest National Laboratory, Richland, WA 99354}                                
\centerline{$^{8}$University of Bern, Bern, Switzerland}                                 
\centerline{$^{9}$The University of Texas at Austin, Austin, TX 78712}             
\centerline{$^{10}$INFN - Laboratori Nazionali del Gran Sasso, Assergi, Italy}                                     
}

\date{\today}

\begin{abstract}
The ArgoNeuT collaboration reports the first measurement of neutral current $\pi^{0}$ production in $\nu_{\mu}$-argon and $\bar{\nu}_{\mu}$-argon scattering. This measurement was performed using the ArgoNeuT liquid argon time projection chamber deployed at Fermilab's NuMI neutrino beam with an exposure corresponding to 1.2$\times 10^{20}$ protons-on-target from the Fermilab Main Injector and a mean energy for $\nu_{\mu}$ of 9.6~GeV and for $\bar{\nu}_{\mu}$ of 3.6~GeV. We compare the measured cross section and kinematic distributions to predictions from the GENIE and NuWro neutrino interaction event generators.

\end{abstract}

\pacs{Valid PACS appear here}
\maketitle


\section{Introduction}\label{sec:Introduction}
Interest in the precise measurement of neutrino-nucleus cross-sections has grown in recent years due to their effect on the interpretation of neutrino oscillation data. Few precise measurements exist for $\nu$ and $\bar{\nu}$ neutral current (NC) neutral pion ($\pi^{0}$) production \cite{PDG, ArgonBubble,DeutData,SingPion,Garg,K2K,MiniBooNECoherent,MiniBooNENCPi0,MiniBooNENCPi02, SciBooNENCPi0,K2KNCPi0}. Cross-section uncertainties for neutral current neutrino scattering become important as precision oscillation measurements attempt to measure charge-parity violation in the neutrino sector ($\delta_{CP}$ mixing parameter)\cite{CPTheory} and disentangle the neutrino mass hierarchy question. The NC$\pi^{0}$ channel is of particular importance to neutrino oscillation experiments as it can be experimentally misidentified as $\nu_{e}$ or $\bar{\nu}_{e}$ charged current production. This misidentification complicates the interpretation of $\nu_{\mu} \rightarrow \nu_{e}$ appearance oscillation measurements, which are required for the detection of neutrino CP violation.

Future short-baseline oscillation experiments, such as MicroBooNE \cite{MicroBooNE}, SBND \cite{LAr1ND}, and ICARUS \cite{ICARUSSBND} as well as long baseline experiments, such as DUNE \cite{LBNE}, plan to utilize large scale liquid argon time projection chambers (LArTPCs)\cite{LArTPC} to detect neutrino interactions. This detector technology offers exemplary electromagnetic shower reconstruction capabilities as well as electron/photon discrimination ability, as recently demonstrated by the ICARUS and ArgoNeuT collaboration \cite{ICARUS,Nu2014}. While a previous measurement of the energy reconstruction of $\pi^{0}$ mesons from cosmic ray production has been performed for LArTPC's \cite{ICARUSPi0}, prior to the measurement presented in this paper, no direct measurement of the neutral current neutrino-argon interaction rate has been performed. The difficulty of identifying NC$\pi^{0}$ interactions in neutrino experiments has lead to few measurements of this process and thus the uncertainty in the cross-section is often a large systematic in $\nu_{e}$ appearance oscillation measurements. The characterization of the NC$\pi^{0}$ production in a LArTPC has increased importance as these LArTPC experiments attempt to disentangle possible hints of new physics from $\nu_{e}$ appearance as reported by the LSND collaboration \cite{LSND} and the MiniBooNE collaboration \cite{MiniBooNE}. 

The unique electron/photon discrimination power offered by LArTPCs will allow future experiments, such as the forthcoming MicroBooNE experiment, to either confirm or rule out any excess seen in electron like events thought to originate from $\nu_{\mu} \rightarrow \nu_{e}$ oscillations. Moreover this $e/\gamma$ discrimination allows LArTPCs the ability to better characterize the dominant background, namely mis-identified $\pi^{0} \rightarrow \gamma \gamma$. In order to accomplish this, precise characterization of NC$\pi^{0}$ production is of the utmost importance.

\begin{figure}[htbp]
  \begin{center}
    \includegraphics[width=0.49\textwidth]{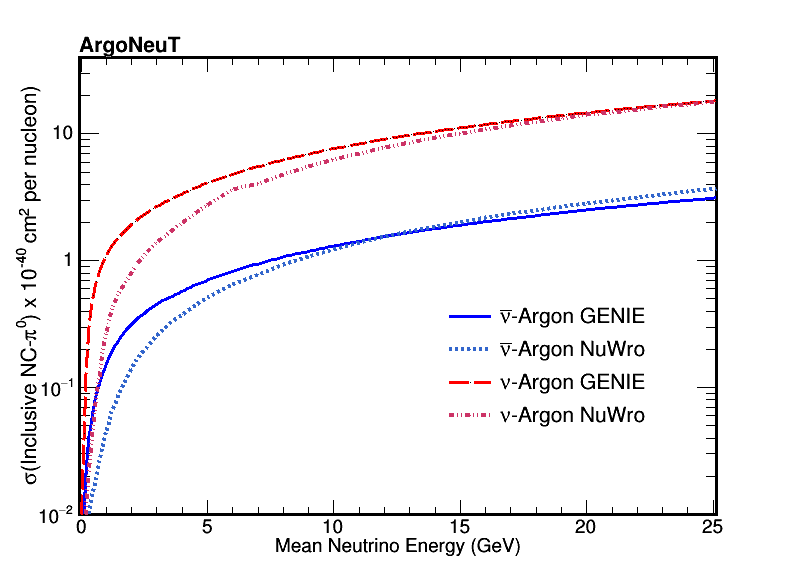}
    \caption{Semi-inclusive neutral current $\pi^{0}$ production as a function of neutrino energy on an argon target as predicted by the GENIE and NuWro event generators.\label{fig:ProductionPrediction}}
  \end{center}
\end{figure}

Fig. \ref{fig:ProductionPrediction} shows the predicted production cross-section for semi-inclusive  NC$\pi^{0}$ utilizing both the GENIE \cite{GENIE} and NuWro \cite{NuWro} neutrino event generators. Both generators predict similar cross-sections in the lower energy region ($<$10~GeV), however the NuWro event generator only includes $\Delta$ resonant production of the $\pi^{0}$ meson, and is thus known to become deficient at higher energies. This known difference does not impact this analysis since the neutrino energies we are interested in are below 10~GeV. Accurate modelling of this production requires knowledge of both the underlying neutrino-nucleon interactions and of final state interactions.

In this paper we present the first measurement of neutrino induced NC$\pi^{0}$ production on an argon target. The interaction final state utilized in this analysis is defined as:
\begin{equation}
\nu_{\mu} + \mbox{Ar} \rightarrow \nu_{\mu} + \pi^{0} + \mbox{X}, 
\end{equation}
\begin{equation}
\bar{\nu}_{\mu} + \mbox{Ar} \rightarrow \bar{\nu}_{\mu} + \pi^{0} + \mbox{X}
\end{equation}
where NC$\pi^{0}$ is defined as an event topology where there is no electron or muon in the final state, at least one $\pi^{0}$ meson that decays to two photons, and any other number of final state nucleons or mesons (X) are present. In the instance where multiple photons are observed in the final state, all possible combinations of photon pairs are considered when attempting to reconstruct the $\pi^{0}$ meson from where the photons originated. This definition differs slightly from much of the historical neutral current $\pi^{0}$ data \cite{ArgonBubble,DeutData,SingPion,Garg,K2K,MiniBooNECoherent,MiniBooNENCPi0,MiniBooNENCPi02, SciBooNENCPi0, K2KNCPi0} which typically require one and only one $\pi^{0}$ meson and little other activity in the detector (typically a single proton). This definition is used in this analysis to help mitigate the low statistics of the data sample. This difference in final state definition, in addition to the neutrino scattering occurring off a much higher Z nuclei such as argon, complicates any direct comparison to historic data. However, where possible comparisons are made to previous measurements.
\section{Overview of the Analysis}\label{sec:AnaOverview}

\begin{figure}[htbp]
  \begin{center}
    \includegraphics[width=0.49\textwidth]{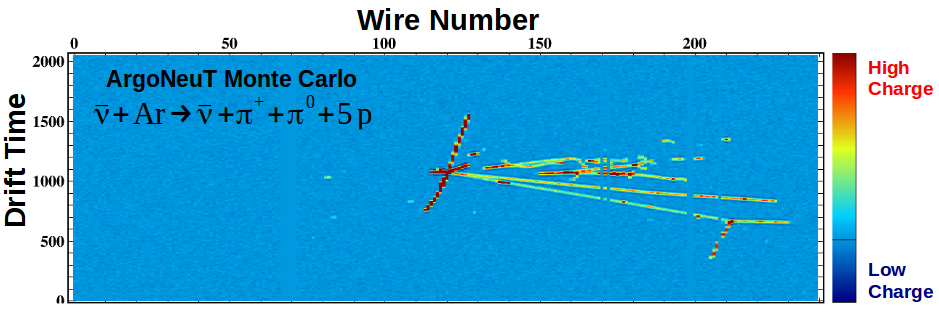}
    \caption{Event display for a Monte Carlo neutral current $\pi^{0}$ event simulated in the ArgoNeuT detector.\label{fig:NCPi0SignalMC}}
  \end{center}
\end{figure}

Fig. \ref{fig:NCPi0SignalMC} shows a simulated Monte Carlo (MC) NC$\pi^{0}$ event inside ArgoNeuT. This particular event demonstrates the semi-inclusive topology in which the neutrino interacts with the nucleus and causes the ejection of a single $\pi^{0}$ meson and a large number of other final state particles. This $\pi^{0}$ then immediately decays into a pair of photons that convert to electron / positron ($e^{+} e^{-}$) pairs a distance characteristic of the 14~cm radiation length of liquid argon from the neutrino interaction point (referred to as the event vertex). The ionization caused by the $e^{+} e^{-}$ pair thus registers on the read-out wires as two clusters of charge (``showers'') pointing back to a common vertex.

In order to identify and reconstruct these NC$\pi^{0}$ events, the analysis proceeds in four parts.
\paragraph*{\textbf{Event Selection}}
First, candidate NC$\pi^{0}$ events are identified utilizing a series of selection criteria that are chosen to reject charge current (CC) interactions and search for topologies consistent with $\pi^{0} \rightarrow \gamma\gamma$ decays inside the ArgoNeuT detector. The selection criteria utilized is outline in Section \ref{sec:EventSelection}.

\paragraph*{\textbf{Energy Corrections}}
ArgoNeuT's small volume causes many of the photons resulting from a $\pi^{0} \rightarrow \gamma\gamma$ decay to not be fully contained within the TPC.  From MC studies, 60$\%$ of the electromagnetic showers coming from $\pi^{0} \rightarrow \gamma\gamma$ decay have less than 50$\%$ of their energy contained. Moreover, 40$\%$ of events have both photon showers with less than 50$\%$ containment. By using the prior that the pair of photons observed in the event come from a decay of a $\pi^{0}$ meson, it is possible to correct back the missing energy due to loss from poor containment based on the opening angles of the photon pair and the topological location of the shower inside the detector. The templates used for the energy corrections are based on the simulation of these events inside the detector as well as the topology of the reconstructed event. MC description of relevant event observables is provided in Section \ref{sec:MCObservables} with the full procedure for the energy corrections and its results given in Section \ref{sec:EnergyCorrections}.

\paragraph*{\textbf{Reconstructed $\pi^{0}$ Kinematics}}
Following the application of the energy corrections, the data is presented as a function of $\pi^{0}$ kinematic variables and the reconstructed photon energy and momentum. 

The momentum of the two photon pair defined as
\begin{equation}\label{eqn:EnergyMomentum}
P_{\gamma\gamma} = \sqrt{E_{1}^{2} + E_{2}^{2}+2E_{1}E_{2} \cos(\theta_{\gamma\gamma})}, 
\end{equation}
where $E_{i}$ is the energy of the photon (ordered by their energy) and $\theta_{\gamma\gamma}$ is the opening angle between the photon pair. Another kinematic observable of the pair of photons is the cosine of the angle of the photon pair ($\gamma\gamma$) with respect to the beam as defined by 
\begin{equation}\label{eqn:CosPi0}
\cos(\gamma\gamma) = \frac{\vec{P}_{z}^{\gamma^{1}} + \vec{P}_{z}^{\gamma^{2}}}{P_{\gamma\gamma}} , 
\end{equation}
where $\vec{P}_{z}^{\gamma^{i}}$ is the z component of the momentum of the photon in the pair and $P_{\gamma\gamma}$ is the reconstructed momentum of the pair of photons (which is the momentum of $\pi^{0}$ mesons in the lab frame if they were correctly identified). Distributions of these quantities may be found in Section \ref{sec:RecoPi0Kin}.

\paragraph*{\textbf{Ratio of NC($\pi^{0}$) to CC}}
One way to interpret the data beyond the reconstructed kinematics of the $\pi^{0}$ and allow a comparison with results obtained from other experiments and theory is to convert the observed event rate into a ratio of efficiency corrected NC$\pi^{0}$ production to efficiency corrected inclusive charged current (CC) production. This ratio can be written as
\begin{equation}\label{eqn:RatioDefn}
Ratio(NC/CC) = \frac{\sigma(NC\pi^{0})}{\sigma(CC)} = \frac{\sum\limits_{NC} \frac{S_{i}^{NC} - B_{i}^{NC}}{\epsilon_{i}^{NC}\Phi_{\nu}N_{Targets}}}{\sum\limits_{CC} \frac{S_{i}^{CC} - B_{i}^{CC}}{\epsilon_{i}^{CC}\Phi N_{Targets}}},
\end{equation}
where $S_{i}^{NC/CC}$ is the number of the signal events in the particular bin from data from NC$\pi^{0}$/CC events, $B_{i}^{NC/CC}$ is the predicted background coming from MC scaled to the appropriate protons-on-target (P.O.T), and $\epsilon_{i}^{NC/CC}$ is the efficiency for NC$\pi^{0}$ or CC events taken from simulation. For identical flux and number of targets Eq. \ref{eqn:RatioDefn} simplifies to
\begin{equation}\label{eqn:RatioSimp}
\frac{\sum\limits_{NC} \frac{S_{i} - B_{i}}{\epsilon_{i}}}{\sum\limits_{CC} \frac{S_{i} - B_{i}}{\epsilon_{i}}} = \frac{N(NC\pi^{0})}{N(CC)}.
\end{equation}
The numerator represents all events with no muon or electron and at least one $\pi^{0}$ observed in the final state exiting the target nucleus. This interaction can be accompanied by any number of other nucleons or other mesons. The denominator represents events with an identified muon in the final state coming from the target nucleus and any other number of other nucleons or final state mesons. One complication in making this simplification arises because the anti-neutrino beam is actually a mixture of a neutrinos and anti-neutrinos. In order to address this, the sample is broken into two components
\begin{equation}\label{eqn:nuRatioOverview}
\frac{N_{\nu}(NC\pi^{0})}{N_{\nu}(CC)} = \frac{\mbox{Number of }\nu \mbox{ induced NC }\pi^{0} \mbox{ Events}}{\mbox{Number of } \nu \mbox{ induced CC Events}}, 
\end{equation}
and 
\begin{equation}\label{eqn:antinuRatioOverview}
\frac{N_{\bar{\nu}}(NC\pi^{0})}{N_{\bar{\nu}}(CC)} = \frac{\mbox{Number of }\bar{\nu} \mbox{ induced NC }\pi^{0} \mbox{ Events}}{\mbox{Number of } \bar{\nu} \mbox{ induced CC Events}},
\end{equation}
ensuring that the simplification made to obtain Eq. \ref{eqn:RatioSimp} takes into account the flux in the anti-neutrino beam due to neutrinos and anti-neutrinos for both the CC and NC sample. In Eq. \ref{eqn:nuRatioOverview} and Eq. \ref{eqn:antinuRatioOverview}, the denominator is taken directly from an analysis of charged current interactions measured in the ArgoNeuT detector \cite{CCInclusive}. This analysis utilized  the MINOS Near Detector (MINOS-ND), a 0.98 kton magnetized steel-scintillator calorimeter \cite{MINOSND}, for non-contained muons exiting in the forward direction to determine the sign of the outgoing lepton and thus could distinguish the species of neutrino interaction as well as the momentum for that lepton. This previous measurement also takes into account the different acceptances for neutrinos and anti-neutrinos.

For the numerator in Eq. \ref{eqn:nuRatioOverview} and Eq. \ref{eqn:antinuRatioOverview}, the neutral current channel, the technique of utilizing the MINOS-ND to distinguish the species of neutrino is not possible. Instead, a MC based estimate on the NuMI beam composition is used to estimate the fraction of NC$\pi^{0}$ events that come from $\nu$ and $\bar{\nu}$ interactions. This means that the number extracted for $N_{\nu}(NC\pi^{0})$ is anti-correlated with $N_{\bar{\nu}}(NC\pi^{0})$ and thus the ratio extracted is also anti-correlated. This separation technique, however, allows for comparisons to other experimental results and thus is a useful tool to interpret the ArgoNeuT data. Details of the procedure to extract this measurement are presented in Section \ref{sec:RatioNCtoCC}.

\paragraph*{\textbf{Flux Averaged NC($\pi^{0}$) Cross-Section}}
Furthermore, we give a measurement of the flux-averaged absolute cross-section for NC$\pi^{0}$ production on an argon nucleus. A similar procedure as described above for extracting the component of the cross-section due to $\nu$ and $\bar{\nu}$ interactions is followed and results are compared to the GENIE and NuWro neutrino event generator in Section \ref{sec:FluxAvgXSection}.

\section{Event Selection}\label{sec:EventSelection}
The ArgoNeuT detector \cite{ArgoNeuT} is a 47.5~$\times$~40~$\times$~90~cm$^{3}$($x$-$y$-$z$) active volume LArTPC with the longest dimension ($z$) situated along the beam axis and two wire planes positioned on beam right. ArgoNeuT ran in the NuMI-LE (Neutrinos at the Main Injector, Low Energy option) beam at Fermi National Accelerator Laboratory \cite{NUMI} and collected 0.085$\times 10^{20}$ protons-on-target (POT) in neutrino mode and 1.20$\times 10^{20}$ POT in anti-neutrino mode. Tab. \ref{tab:Fluxes}, taken from Ref. \cite{CCInclusive}, provides the total flux for the anti-neutrino mode beam used in this analysis. 

\begin{center}
\begin{table}[htb]
	\begin{center}
	\begin{tabular}{c|c|c}
	\multicolumn{3}{c}{\textbf{Total anti-neutrino mode fluxes}} \\
	\hline \hline
	 \textbf{$E_{\nu}$} GeV & $\nu$ Flux & $\bar{\nu}$ Flux  \\
	 0 - 50 & 3.9 $\pm 0.4 \times 10^{7}$ & 2.4 $\pm 0.3 \times 10^{5}$ \\
	\hline
	\end{tabular}
	\caption{The neutrino and antineutrino fluxes for the anti-neutrino mode beam, taken from Ref. \cite{CCInclusive}, used in this analysis. The flux unit is $\nu_{\mu}$/GeV/m$^{2}$/$10^{9}$ POT} \label{tab:Fluxes}
	\end{center}
\end{table}
\end{center}

A 481 V/cm electric field is imposed that allows ionization trails created by charged particles traversing the argon medium to be drifted toward the sensing wires. The signal from the wire planes, oriented 60$^{\circ}$ with respect to one another, is combined to provide 3D reconstruction of the neutrino interaction along with particle identification and calorimetric information. 

ArgoNeuT utilizes the LArSoft software package \cite{larsoft} that provides a full rendering of charged particles interacting inside the ArgoNeuT detector. LArSoft provides a full simulation of the experiment and electronics response as well as a simulation of neutrino interactions utilizing the GENIE neutrino event generator and GEANT4 \cite{geant4} for the propagation of particles inside the detector. The propagation of particles into the MINOS-ND is done using GEANT3 \cite{geant3} and a standalone version of the MINOS-ND simulation is employed to characterize the matching of tracks passing from ArgoNeuT to MINOS. Monte Carlo events are treated in the reconstruction package identically as data events.

Three volumes inside the ArgoNeuT detector are defined and used throughout this analysis. All three volumes employ a right-handed coordinate system with positive $Y$ vertical, positive $Z$ parallel to the neutrino beam axis, and the origin placed at the upstream end of the ArgoNeuT TPC, with the detector centered on $Y=0$, and $X=0$ at the TPC sense-wire plane. We also use the conventional polar angles $\theta$ and $\phi$ to denote vector directions, as well as $\theta_x$ and $\theta_y$, the angles with respect to the $X$ and $Y$ axes, respectively.
\begin{itemize}
\item \textbf{Active Volume:}
Is the volume of the entire ArgoNeuT TPC defined as: 0~cm$<$~$X$~$<$47.5~cm, -~20~cm$<$~$Y$~$<$20~cm, 0~cm$<$~$Z$~$<$90~cm. 
\item \textbf{Fiducial Volume:}
A volume definded to allow for a small volume of argon between any interaction and the active boundary. This distance is chosen to mirror that used in the inclusive charged current analysis. This allows for a ratio between the measured neutral current and charge current rate to be easily compared. The fiducial volume is thus defined as: 3~cm$<$~$X$~$<$44.5~cm, -~16~cm$<$~$Y$~$<$16~cm, 6~cm$<$~$Z$~$<$86~cm. 
\item \textbf{Photon Conversion Volume (PCV):}
A volume defined such that a volume of argon exists between the point where the photon converts to an $e^{+} e^{-}$ pair (defined as the photon vertex) and the boundary of the detector. This volume allows for a photon that converts near the boundary to still be identified via a dE/dX measurement. The photon conversion volume is defined as: 5~cm$<$~$X$~$<$42.5~cm, -~15~cm$<$~$Y$~$<$15~cm, 5~cm$<$~$Z$~$<$85~cm.
\end{itemize}

The signal events for this analysis are characterized by a neutrino interaction occurring inside the active volume of the detector that produces at least one $\pi^{0}$ via a neutral current interaction and subsequently decays to a pair of photons. In order to be considered a neutral current interaction no track may be reconstructed and identified as a muon or electron of either sign. Photons from the decay of the $\pi^{0}$ must convert to $e^{+} e^{-}$ pairs inside the PCV in order to be considered in our selection. 

Background events for this analysis are categorized inclusively as any event that is not already identified as NC$\pi^{0}$ event. Four selection requirements are used to identify candidate NC$\pi^{0}$ events. These selection requirements are chosen to reject events that appear to come from a charged current interaction, and thus produce a charged lepton in the detector volume, and to identify events that have a topology consistent with the presence of a $\pi^{0} \rightarrow \gamma\gamma$ decay.

We first reject events in which a muon track found in MINOS-ND that is matched to a track in the ArgoNeuT detector. The front face of MINOS-ND is approximately 1.5 m downstream of ArgoNeuT, and the center of ArgoNeuT is located 20 cm below the center of the MINOS fiducial volume. An ArgoNeuT-MINOS-ND track match is defined by the following criteria: (i) the track has a MINOS-ND hit within 20~cm of the front face of the MINOS-ND detector; (ii) the MINOS-ND track must start within 35~cm of the projected ArgoNeuT track location in the $y-z$ plane; (iii) the ArgoNeuT and MINOS-ND track direction cosine differences must satisfy the requirements $|\delta\cos(\theta_{x})|<1.0$, $|\delta\cos(\theta_{y})|<1.0$, and $|\delta\cos(\theta_{z})|<0.5$. These selection requirements are similar (although much more inclusive) to those used in previous ArgoNeuT CC analyses (\cite{CCInclusive, Coherent}) and have been shown to be efficient at identifying charged current interactions within the ArgoNeuT detector. If such a track exists, the event is rejected as likely coming from a charged current interaction. The dominant inefficiency for this selection comes from the incorrect matching of non-related tracks in the MINOS-ND to charged pion tracks present in the NC$\pi^{0}$ interaction.


The next selection requirement applied is designed to reject charged current events missed by the anti-matching to the MINOS-ND. These events can fail the anti-matching because the muon produced in a charged current interaction does not exit in the direction of the MINOS-ND or because the track is poorly reconstructed and thus does not match back to the track found in MINOS-ND. To reject these events we utilize the reconstruction information from inside the ArgoNeuT TPC and veto an event that has a topology consistent with having a muon originating from a neutrino interaction vertex, as expected in a charged current interaction. These events must have at least two tracks that are identified as either a muon/proton, muon/pion or muon/muon pair emanating from a common vertex. This selection was chosen over simply removing any event with a minimum ionizing track (MIP) present in the event in order to preserve the statistics of the selected sample and avoid removing neutral current events with charged pions in the final state that are incorrectly identified. The dominant inefficiency comes from multiple misidentification of charged pions as muons thus causing NC$\pi^{0}$ events to be incorrectly excluded from the sample. However, this is a relatively small inefficiency compared to other selection criteria. The choice to not reject any event with a minimum ionizing track was made to help increase the statistics of the sample of candidate NC$\pi^{0}$ events where, in addition to the neutral pion, charged pions and high momentum protons which exit the chamber are present.  

\begin{figure}[htbp]
  \begin{center}
    \includegraphics[width=0.50\textwidth]{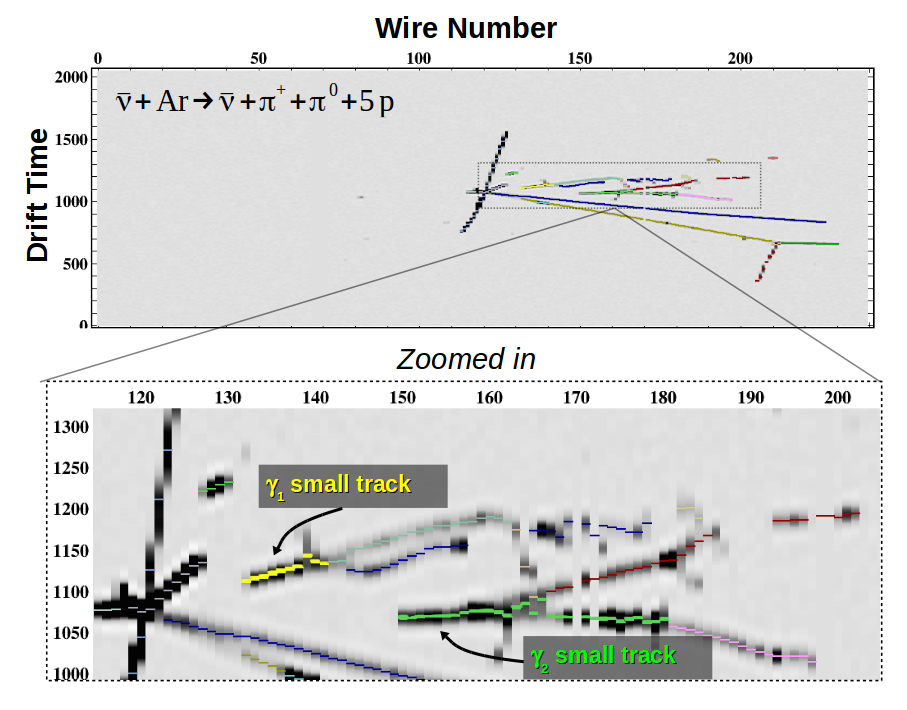}
    \caption{A simulated neutral current $\pi^{0}$ event reconstructed using the small track algorithm in order to break up an electromagnetic shower into smaller track like segments for analysis and identification. Inside the photon showers, highlighted in the image, the unique colors indicate the individual small tracks that have been reconstructed.  \label{fig:PhoTrackletExample}}
  \end{center}
\end{figure}

The next event selection identifies $\pi^{0} \rightarrow \gamma\gamma$ decays by leveraging the powerful track reconstruction techniques available in LArTPC's. Correlated groups of short-length tracks consistent with electrons or positrons produced in electromagnetic showers can be identified and reliably separated from tracks produced by pions, muons, and protons. The selection accomplishes this identification by analyzing the components of an electromagnetic shower as if it is made of many small tracks and attempting to identify the starting ``trunk'' of the shower. The electromagnetic nature of short tracks can be verified by a particle identification (PID) procedure that correlates energy loss in the TPC, $dE/dX$, with the range of the tracks in liquid argon, as well as taking into account the topological reconstruction of the small tracks. This also allows for a determination of the energy of each track. Fig. \ref{fig:PhoTrackletExample} shows a simulated event where the electromagnetic showers have been broken into smaller components based on their track-like structure. Importantly, the beginning of the shower is reliably reconstructed as a small track component of the shower and can be identified by analyzing the reconstructed components.

The procedure for selecting $\pi^{0} \rightarrow \gamma\gamma$ topologies using these small tracks is to first require that at least two such small tracks are found in the event. If during the application of any subsequent requirement the number of small tracks present is less than two, the event is removed from consideration. Short tracks with a PID assignment consistent with that of a muon, pion, or proton are removed from consideration. Next, pairs of small tracks are kept for consideration if their start points are separated by at least 4.0~cm. This requirement identifies pairs of small tracks coming from a pair of photons that are separated in space removing highly ionizing parts of a single shower that have been broken into many small tracks while preserving the unique starting portion of the distinct shower pairs. Next we require greater than 75~$\%$ of the first 4~cm of the track to have a $dE/dX$ value $\geq$~3.5~MeV/cm. If the track is shorter than 4~cm in length than the requirement is that the majority ($\geq50\%$) track have a $dE/dX$ $\geq$~3.5~MeV/cm. This selection is designed to only keep highly ionizing short tracks, such as those coming from a photon conversion into an $e^{+} e^{-}$ pairs, and reject those likely due to a minimum ionizing particle. Finally, to separate distinct photon conversions from the background of electron and single photon, we cluster the hits in the event using a density based spatial clustering algorithm (DBSCAN) and require that any small track share less than 85$\%$ of its hits with any other small track from within the same DBSCAN based cluster. This DBSCAN algorithm groups together nearby distributions of charge and associates them into one object. The result of this requirement is that two spatially close small tracks that belong to the same DBSCAN cluster (sharing the majority of their hits) will be removed from consideration.

Finally, events are visually examined (``hand-scanned'') by physicists (``scanners'') to identify two electromagnetic showers originating from a NC$\pi^{0}$ interaction. The hand-scan procedure both identifies events consistent with a $\pi^{0} \rightarrow \gamma\gamma$ decay and rejects background events that have passed prior selection requirements. In general, the hand-scan of events takes place in three parts. Step one selects events with associated photon showers resulting from a NC $\pi^{0}$ interaction. Here the topology of the event is taken into consideration and the scanner looks for two showers pointing back to a common vertex point. Part two has the scanner select the clusters in each view by defining the shower's start point and axis. The shower axis represents a cylinder around which hits associated with the shower will be selected. When the start point and axis are identified for both planes of wires, the preliminary 3D angle shower object is created. In step three the scanner includes or excludes any hits not initially associated to the shower and builds the 3D shower object and evaluates the quality of the reconstructed shower.

The scanner has the ability to review the shower selection by reconstructing the 3D shower's vertex location, angles ($\theta , \phi$), reconstructed energy, and $dE/dX$ over the first 2.4 cm of the shower. This allows the scanner to check that the shower is consistent with a pair converting photon (dE/dX values $\geq$ 2.5 MeV/cm) and begins within the appropriate boundary (PCV). If the shower is deemed to have been correctly identified, it is then processed into an offline file for further analysis. Once fully reconstructed offline, showers whose $dE/dX$ profile in the first 2.4~cm of the shower is greater than 2.5 MeV/cm, to distinguish photon from electron induced showers, are kept for analysis.

\section{Matching Electromagnetic Shower Angles}\label{sec:MCObservables}
\begin{figure}[htb]
  \begin{center}
    \includegraphics[width=0.49\textwidth]{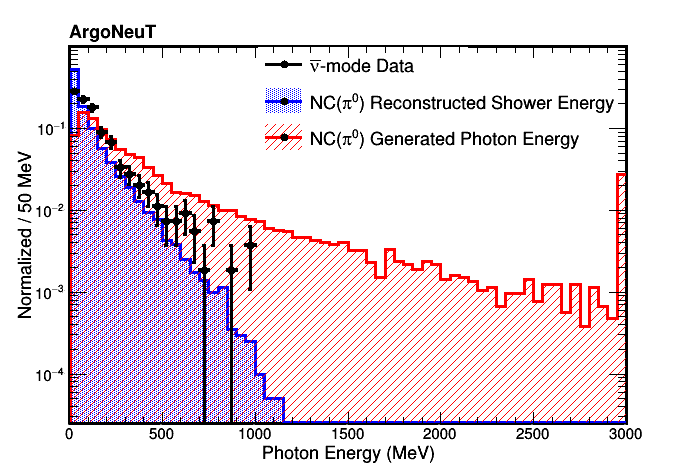}
    \caption{The reconstructed energy of candidate  NC$\pi^{0}$ events compared to the simulated deposited energy from  NC$\pi^{0}$ interactions inside the active volume (blue) and the MC true energy of the photons (red). The distributions have been area normalized.} \label{fig:CompareEnergy} 
 \end{center}
\end{figure}

Following the identification of candidate NC$\pi^{0}$ events, the reconstruction of the selected electromagnetic showers becomes necessary in order to further identify events consistent with coming from $\pi^{0} \rightarrow \gamma\gamma$ decay. In order to do this, the energy of the shower must be obtained. However, as will be discussed further in Section \ref{sec:EnergyCorrections}, the majority of photon showers coming from $\pi^{0}$ decays are not contained within the ArgoNeuT detector. Fig. \ref{fig:CompareEnergy} demonstrates the problem by plotting the reconstructed energy of the candidate data events compared to the true MC Energy for NC$\pi^{0}$ events and the simulated deposited energy inside the active volume of the detector. 

\begin{center}
\begin{figure}[htb]
  \begin{center}
    \includegraphics[width=0.49\textwidth]{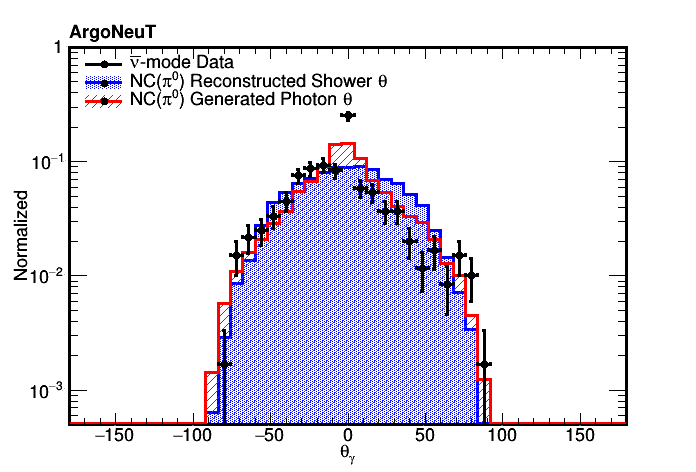}
    \includegraphics[width=0.49\textwidth]{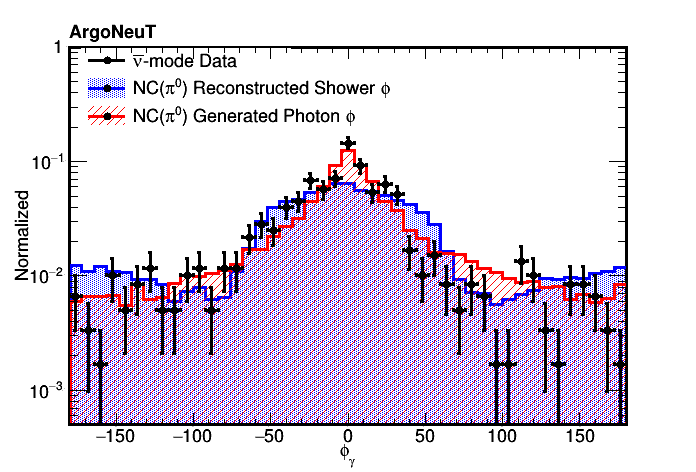}
    \caption{The reconstructed angles theta ($\theta$) and phi ($\phi$) for the photon showers of candidate  NC$\pi^{0}$ events compared to the simulated deposited angle from  NC$\pi^{0}$ interactions inside the active volume (blue) and the Monte Carlo true angles of the photons (red). The deposited angles are calculated using the start point of the shower and a charge weighted sum of the visible shower.   \label{fig:CompareAngles}} 
 \end{center}
\end{figure}
\end{center}

However, utilizing the fine grain tracking detection of LArTPC's, it is possible to reconstruct the angle of the photons given the visible portion of the shower in the detector. Fig. \ref{fig:CompareAngles} shows the distribution for the candidate NC$\pi^{0}$ data events compared to the simulated deposited angle calculated using the charge weighting of the visible shower and the MC true angle. Using the reconstructed start point and angle of the shower, made possible by the tracking capabilities of the detector, the data closely tracks the true angles. 


\begin{figure}[htb]
  \begin{center}
    \includegraphics[width=0.49\textwidth]{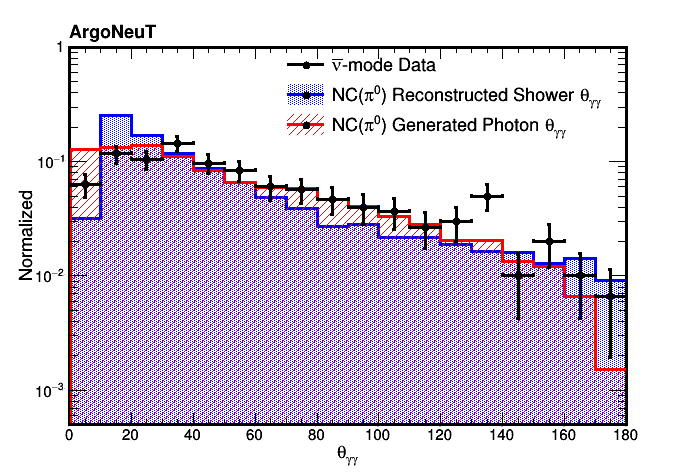}
    \caption{The reconstructed opening angle between two electromagnetic showers for candidate NC$\pi^{0}$ events compared to the simulated opening angle from NC$\pi^{0}$ interactions inside the active volume (blue) and the simulated true opening angle of the photons (red). The distributions have been area normalized.} \label{fig:CompareThetaGG}
 \end{center}
\end{figure}

The well-reconstructed shower angles permit extraction of information about the energy and momentum of the event by utilizing the assumption that the two electromagnetic showers originate from the decay of $\pi^{0} \rightarrow \gamma\gamma$ and constructing the opening angle between the two photons ($\theta_{\gamma\gamma}$). Fig. \ref{fig:CompareThetaGG} shows the $\theta_{\gamma\gamma}$ distribution for the candidate NC$\pi^{0}$ data events compared to the simulated deposited angle that would be calculated using only a charge weighting of the visible shower and the simulated true angle.  The distribution tracks well with the true information up to small opening angle, where the reconstruction has difficulty disentangling the two showers. This provides confidence to the hypothesis that these events do in fact come from a $\pi^{0} \rightarrow \gamma\gamma$ and we will utilize this ability to reconstruct angles well within a LArTPC in the next section to attempt to correct back the missing energy. 

\section{Energy Corrections}\label{sec:EnergyCorrections}
\begin{figure}[htb]
  \begin{center}
    \includegraphics[width=0.49\textwidth]{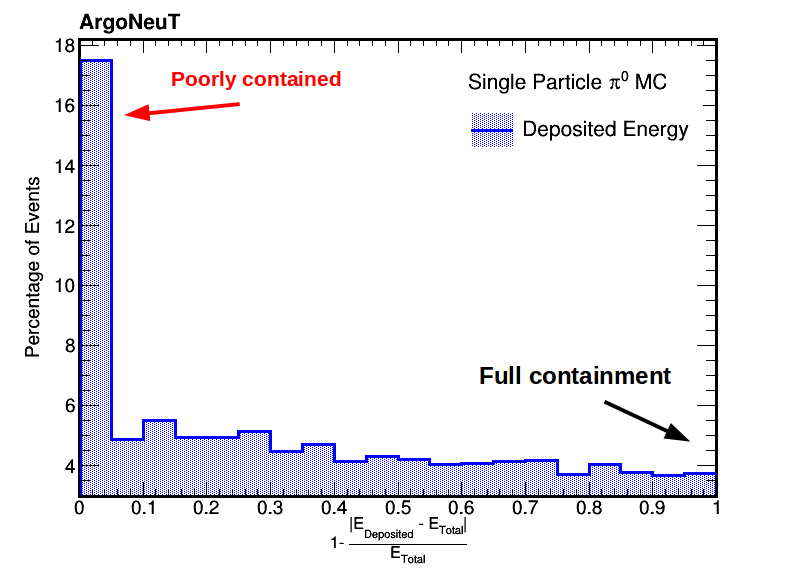}
    \caption{The fraction of energy from simulated $\pi^{0} \rightarrow \gamma\gamma$ events contained in the ArgoNeuT volume. These events have a mean momentum of 0.5 GeV and are simulated uniformly in their initial ($x$, $y$, $z$) location in the TPC volume.} \label{fig:EnergyDeposited}
 \end{center}
\end{figure}

Having reconstructed the portion of the shower that is contained inside ArgoNeuT, corrections must be developed to model the portion of the shower that is not contained within the active volume. The mean momentum for $\pi^{0}$'s created in a neutrino-argon interaction in the ArgoNeuT detector is 730 MeV. At these energies, the photon showers have a radiation length of 14~cm and photons typically convert to electron positron pairs within three radiation lengths, which is the same order as the size of the volume of the ArgoNeuT TPC. Furthermore, these $\pi^{0}$'s can be created anywhere within the fiducial volume, thus the probability of the energy containment of both photons is low. Fig. \ref{fig:EnergyDeposited} shows the simulated energy containment from photons coming from $\pi^{0}$'s with a similar momentum and position distribution as the NC$\pi^{0}$ sample. The fraction of the energy contained is defined as
\begin{equation}\label{eqn:EnergyDep}
1 - \frac{|E_{Deposited} - E_{Total}|}{E_{Total}},
\end{equation}
where $E_{Deposited}$ is the total energy deposited by the electromagnetic shower from the photon inside the active volume. $E_{Total}$ is the true energy of the photon that caused the electromagnetic shower. This definition is chosen such that the energy containment of the shower is between zero (poor containment) and one (full containment)

The dominant cause of energy loss is the shower exiting the boundary of the detector. This happens when a photon converts near the boundary and is directed toward one of the TPC walls, thus resulting in the majority of its energy escaping. Given the radiation length of a photon in argon, the photon would have to convert very far away from any boundary of the ArgoNeuT TPC in order to have near full containment. 

A series of corrections based on the topological reconstruction of the visible component of the electromagnetic showers are applied. These corrections aim at adding back the energy loss due to poor containment of the electromagnetic shower inside the active volume.

The details of the derivation of the energy corrections are left for discussion in Appendix \ref{sec:EnergyCorrAppend}, but broadly speaking the approach adopted for this analysis corrects back this energy loss taking as a prior that two identified photon showers from the event selection described in Section \ref{sec:EventSelection} come from the decay of a $\pi^{0}$. Utilizing this assumption, a relationship between the opening angle of the two photons, $\theta_{\gamma\gamma}$, and the momentum of the $\pi^{0}$, $P_{\pi^{0}}$, is derived. This relationship provides the basis for the application of the subsequent energy corrections. Specifically, any energy correction that is applied to a photon based on its distance to the nearest boundary is not allowed to cause the calculated $P_{\gamma\gamma}$ from Eq. \ref{eqn:EnergyMomentum} to exceed the inferred momentum from $\theta_{\gamma\gamma}$. The amount of containment of the electromagnetic shower within the active volume of the detector is determined by simulating a large number of $\pi^{0} \rightarrow \gamma\gamma$ events inside the ArgoNeuT TPC and building templates for the characteristics of the energy loss based on the location of the photon's conversion. The procedure for constructing the templates is broken into three steps described in greater detail next.

\subsection{Angle hypothesis of the $\pi^{0}$ momentum}
\begin{figure}[htb]
  \begin{center}
    \includegraphics[width=0.48\textwidth]{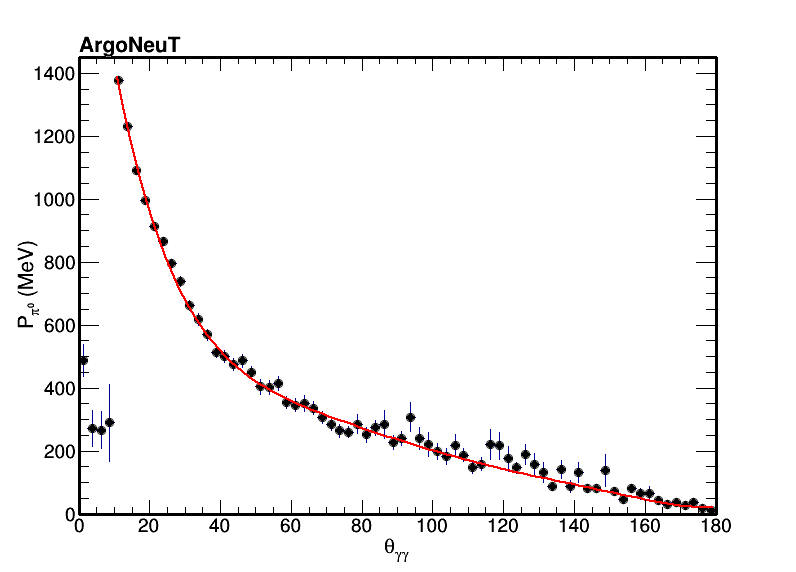}
    \caption{The projection of the $\pi^{0}$ momentum versus the mean opening angle between the two decay photons. The full 2-d relationship is given in Fig. \ref{fig:MomentumVsAngle2d} in Appendix \ref{sec:AngleAppend}. The fitted function provides a bound on the momentum of the hypothesized $\pi^{0}$ system, which any subsequent energy correction to the photons is not allowed to violate. \label{fig:MomentumVsAngle}}
 \end{center}
\end{figure}

An event that is identified as having two or more reconstructed showers will have a hypothesis formed assuming that two of the showers come from the decay of a $\pi^{0}$. Using this hypothesis, an initial estimate of the momentum of the $\pi^{0}$ ($P_{\gamma\gamma}$) responsible for these two electromagnetic showers is calculated based on the opening angle between the showers. When the $\pi^{0}$ decays the photons have an angle $\theta_{\gamma\gamma}$~$\leq$180$^{\circ}$ due to the boost from the rest frame to the lab frame. The greater the momentum of the $\pi^{0}$ in the lab frame the smaller the opening angle $\theta_{\gamma\gamma}$ will be, as shown in Fig. \ref{fig:MomentumVsAngle}. This correlation breaks down for angles less than 10$^{\circ}$ due to the difficulty of disentangling overlapping showers.  To derive the correlation between $\theta_{\gamma\gamma}$ and $P_{\pi^{0}}$ (between 10$^{\circ}$ and 180$^{\circ}$) a polynomial fit is used to provide an analytical expression
\begin{equation}\label{eqn:AngleMomentum}
P_{\gamma\gamma} = C_{0} + \sum_{i = 1}^{6} C_{i}\theta_{\gamma\gamma}^{i},
\end{equation}
with the constants $C_{i}$ given in Appendix \ref{sec:AngleAppend}. 

Comparing the $P_{\pi^{0}}$ calculated using the energy of the photon showers deposited in the active volume given in Eq. \ref{eqn:EnergyMomentum} to the polynomial fit to the hypothesized $P_{\pi^{0}}$ obtained from the opening angle distribution allows a bound to any subsequently applied energy correction. Namely, when considering the application of a correction to one of the two photons ($\gamma_{1}, \gamma_{2}$), due to its containment, the subsequently calculated $P_{\pi^{0}}$ using Eq. \ref{eqn:EnergyMomentum} must not exceed the $P_{\pi^{0}}$ calculated using Eq. \ref{eqn:AngleMomentum}. The opening angle sets a minimum value on $P_{\pi^{0}}$ from the kinematics. However, the distribution in opening angle becomes rather sharply peaked about the minimum as $P_{\pi^{0}}$ increases. So, while not excluded from kinematics, a value higher than $P_{\pi^{0}}$ from opening angle is unlikely, and becomes less likely as the opening angle decreases.

Thus with each energy correction applied to one of the photons, one can evaluate if this correction would exceed a reasonable hypothesis for the momentum of the $\pi^{0} \rightarrow \gamma\gamma$ system. If it would, the correction is not applied to the photon. If it does not, the correction is applied and the next energy correction is attempted. The detailed procedure for applying this hypothesis is provided in Appendix \ref{sec:AngleAppend}.

\subsection{Energy Corrections}

With the momentum hypothesis, $P_{\gamma \gamma}$, formed a series of energy corrections are attempted to correct back the energy loss due to poor containment within the ArgoNeuT TPC. The details of the corrections are described in the Appendix in sections  \ref{sec:LinearCorrAppend} and \ref{sec:XYZAppend}, an overview of which we cover here. 

Photons of from higher momentum $\pi^{0}$'s are less well contained within the TPC and thus are subject to having their energy missed due to containment. Using Eq. \ref{eqn:AngleMomentum} to estimate the initial momentum of the $\pi^{0}$ system an energy correction is applied to $\gamma_{1}$ and $\gamma_{2}$ (where $\gamma_{1}$ is ordered such that it is the most energetic of the photon pair). $P_{\pi^{0}}$ is then calculated using Eq. \ref{eqn:EnergyMomentum} and if the correction is found to violate the initial hypothesis formed using the opening angle, the correction is not applied to $\gamma_{1}$ but instead applied to $\gamma_{2}$ and $P_{\pi^{0}}$ is recalculated. If after this, the correction is still found to violate the initial hypothesis the correction for $\gamma_{1}$ is swapped for the correction to $\gamma_{2}$ and the procedure is repeated. The energy correction can be applied to both, either, or none of the photons for any given event.

Next, a set of corrections are applied based on where the shower vertex is located inside the detector and the direction the shower is pointing. The amount of energy that is deposited inside the detector is strongly correlated with where the photon first converts inside the TPC and how much argon there is between the vertex and the nearest TPC boundary. The basis of these energy corrections depends on the well known electromagnetic shower profile within liquid argon \cite{SlacLArEM,FNALLArEM}. We calculate the distance to the nearest wall in $x$, $y$, $z$ space using the ``straight line'' distance between the photon shower vertex and the nearest boundary which the shower is pointed towards. To correct back the energy loss due to this topological location of the electromagnetic shower, we plot the fraction of energy contained (as defined in Eq. \ref{eqn:EnergyDep}) versus the distance to the boundary. A polynomial fit provides a functional form for the energy correction based on its distance to that given boundary. Similar to before, each of these corrections must not violate the initial $P_{\pi^{0}}$ hypothesis formed using Eq. \ref{eqn:AngleMomentum}, and a correction is kept for any individual photon only if these criteria are met.

\subsection{Results of Energy Corrections}\label{sec:EnergyCorr}
\begin{figure}[htb]
  \begin{center}
    \includegraphics[width=0.48\textwidth]{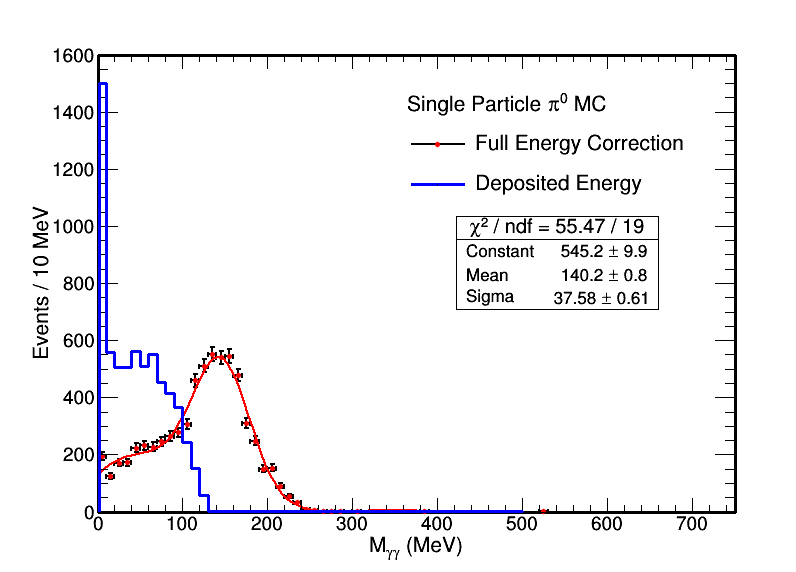}
    \caption{The invariant mass of the $\gamma\gamma$ system after the application of the template based energy corrections for a MC sample of $\pi^{0}$ decays inside the ArgoNeuT detector. The distribution is fit with a Gaussian plus polynomial between 0~MeV$\leq M_{\gamma\gamma}\leq$240~MeV. This fit returns a mean value of 140.2$\pm$0.8~MeV with a RMS of 37.6$\pm$0.61~MeV. } \label{fig:InvMassFinal}
 \end{center}
\end{figure}

The complete set of results utilizing the application of the template based energy corrections and the procedure described briefly above is given in Appendix \ref{sec:ECorrResultsAppend}. From MC studies, less than 10$\%$ of events receive energy corrections that would move the observed shower energy above the true value. Fig. \ref{fig:InvMassFinal} shows the outcome of the full suite of energy corrections to the reconstructed $\gamma\gamma$ invariant mass, defined as $M_{\gamma\gamma} = \sqrt{4 E_{1}^{\gamma}E_{2}^{\gamma} \sin^{2}(\frac{\theta_{\gamma\gamma}}{2})}.$

The distribution is fit to a Gaussian plus a polynomial (to model the low energy mis-reconstruction) between 0~MeV$\leq M_{\gamma\gamma}\leq$240~MeV. The result of the fit returns a mean value for the Gaussian of 140.2$\pm$0.8~MeV with a width of 37.5$\pm$0.61~MeV (to be compared to $M_{\pi^{0}} = 135$~ MeV). This demonstrates that the template based energy corrections do adjust back the photon's energy closer to its true energy and thus give us another tool to identify candidate NC$\pi^{0}$ events.



\section{Reconstructed $\pi^{0}$ Kinematics}\label{sec:RecoPi0Kin}
After applying all corrections to the energy of the showers in our data sample, we require the reconstructed invariant mass, $ M_{\gamma\gamma}$, to lie within the range 60~MeV$\leq M_{\gamma\gamma}\leq$240~MeV. This requirement selects events that reconstruct inside $\pm$2$\sigma$ of the invariant mass of the $\pi^{0}$. The size of the window is selected based on the reconstructed mass peak RMS from the sample of events used to calibrate the energy of the photons. 

Fig. \ref{fig:DataInvMassPlot} shows the reconstructed invariant mass distribution of anti-neutrino mode data events before the requirement that all events fall between 60~MeV$\leq M_{\gamma\gamma}\leq$240~MeV. A Gaussian plus linear function is fitted to the data, yielding an invariant mass of $131.1 \pm 8.4$ MeV with a width of $81.4 \pm 11.1$ MeV. After requiring events with an invariant mass between 60~MeV$\leq M_{\gamma\gamma}\leq$240~MeV, we reconstruct the invariant mass with a peak of $138.4\pm7.0$ MeV and a width of $54.7\pm7.2$ MeV. The fitted mean is consistent with the 135 MeV $\pi^{0}$ mass taking into account statistical uncertainties and the systematic energy scale error associated with the energy correction scheme.

\begin{widetext}
\begin{center}
\begin{figure}[htb]
  \begin{center}
    \includegraphics[width=0.48\textwidth]{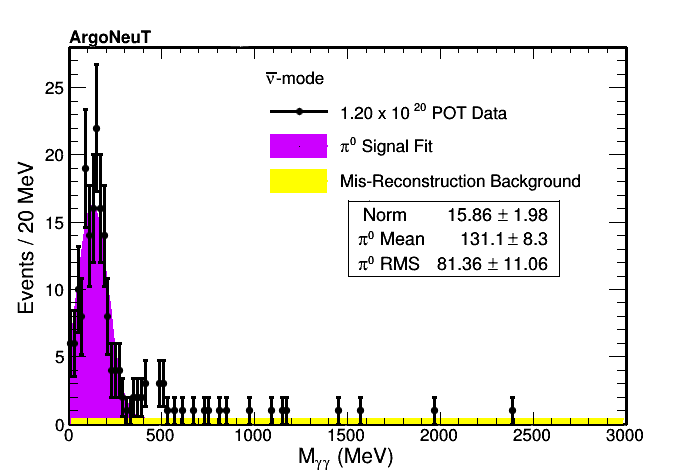}
    \includegraphics[width=0.48\textwidth]{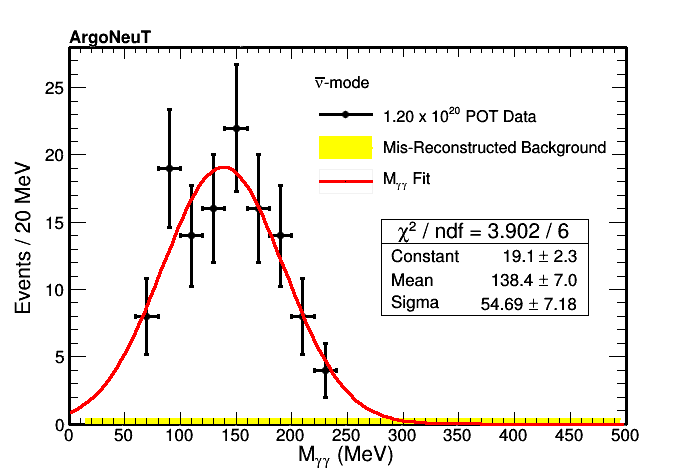}
    \caption{(Left) $\pi^{0}$ invariant mass plot for all NC$\pi^{0}$ candidate events before requiring that all events fall between 60~MeV$\leq M_{\gamma\gamma}\leq$240~MeV. A clear peak near the invariant mass of the $\pi^{0}$ meson can be seen. (Right) $\pi^{0}$ invariant mass plot zoomed into the invariant mass peak between 60~MeV$\leq M_{\gamma\gamma}\leq$240~MeV. The mean of the fitted Gaussian shifts after the selection of events between 60~MeV$\leq M_{\gamma\gamma}\leq$240~MeV because the low invariant data points have been excluded from the fit. These events include interactions from both $\nu$ and $\bar{\nu}$ scattering events. \label{fig:DataInvMassPlot}}
 \end{center}
\end{figure}
\end{center}
\end{widetext}

Tab. \ref{tab:CutSummary} summarizes the effects of all selection criteria and the application of the energy correction procedure used for this analysis. In total, 123 data events survive all the cuts, consistent with the expected 159 events (101 NC$\pi^{0}$ events and 58 background events) predicted from the GENIE MC. As evident from Fig. \ref{fig:DataInvMassPlot}, the data are consistent with a model where $\pi^{0}$ production fully accounts for the two photon mass distribution in the range 60-240 MeV. This observation is consistent with predictions from the MC using the GENIE production model defined in Appendix B.

\begin{widetext}

\begin{table}[htb]
	\begin{center}
	\begin{tabular}{|c|c|c|c|}
	\hline
	 & \textbf{Number of MC Events} &  &\\
	  \textbf{Event Selection} & \textbf{Scaled to 1.20$\times10^{20}$ POT} & \textbf{Signal Acceptance} & \textbf{Background Rejection}  \\
	  & (Signal / Background) & $\%$ & $\%$ \\
	\hline
	Total Number of Events & 615 / 10,019 & - & - \\
	\hline
	Anti-MINOS Matching & 494 / 2,475& 80$\%$ &75$\%$ \\
	\hline
	Charged Current Veto& 365 / 1,664 & 74$\%$ & 33$\%$  \\
	\hline
    Small Track Reconstruction& 285 / 792 & 78$\%$ & 52$\%$ \\
	\hline
	Shower Reconstruction & 188 / 126 & 66$\%$ & 84$\%$ \\
	\hline
	$<$dE/dX$>$ $\geq$ 2.5 MeV/cm & 158 / 107 & 84$\%$ & 15$\%$  \\
	\hline
	60~MeV$\leq M_{\gamma\gamma}\leq$240~MeV & 101 / 58 & 64$\%$ & 46$\%$  \\
	\hline
	\hline
	\textbf{Data Passing all event selection} & 123 &\multicolumn{2}{c|}{ } \\
	\hline
	\end{tabular}
	\caption{Summary of NC $\pi^{0}$ event selection cuts applied to ArgoNeuT Monte Carlo and data.} \label{tab:CutSummary}
	\end{center}
\end{table}

\end{widetext}

Of the background events remaining after all the cuts are applied, 95.7$\%$ are charged current events with a $\pi^{0}$ produced in the neutrino interaction. These are events where the muon was not reconstructed well enough to be matched to the MINOS-ND nor be identified within the TPC. The remaining 4.3$\%$ results from either a mis-identified particle as a photon shower or a $\pi^{+}$ created in the neutrino interaction that underwent charge exchange producing a $\pi^{0}$ that is mis-identified as coming from the primary interaction point. 

Fig. \ref{fig:PiDataPlots} shows the momentum of the $\pi^{0}$, from Eq. \ref{eqn:EnergyMomentum} and the cosine of the angle of the $\pi^{0}$ with respect to the beam as defined by Eq. \ref{eqn:CosPi0}. Despite the low statistics, both of these distributions have general agreement with the MC prediction in shape while the MC over predicts the peak. 

\begin{widetext}
\begin{center}
\begin{figure}[htb]
  \begin{center}
    \includegraphics[scale=0.36]{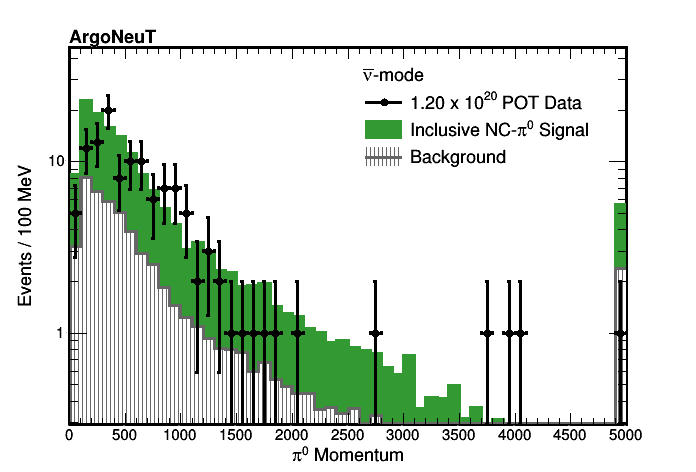}
    \includegraphics[scale=0.36]{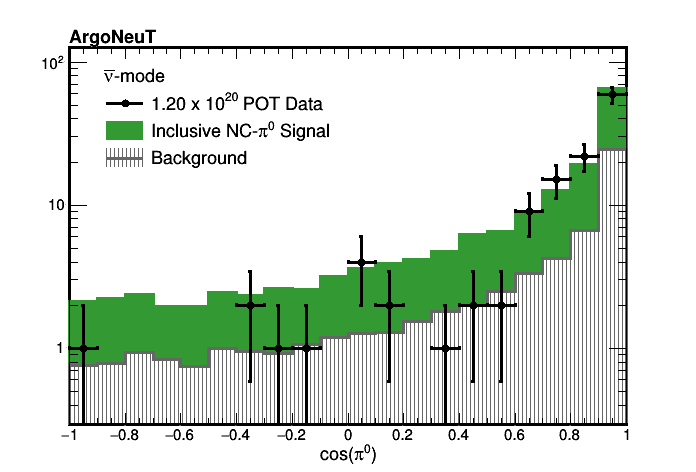}
    \caption{$\pi^{0}$ Momentum and cosine of the angle of the $\pi^{0}$ for the set of anti-neutrino data events with MC backgrounds scaled to 1.2$\times 10^{20}$ POT. These events include interactions from both $\nu$ and $\bar{\nu}$ scattering events. \label{fig:PiDataPlots}}
 \end{center}
\end{figure}
\end{center}
\end{widetext}


\begin{widetext}
\begin{center}
\begin{figure}[htb]
  \begin{center}
    \includegraphics[scale=0.35]{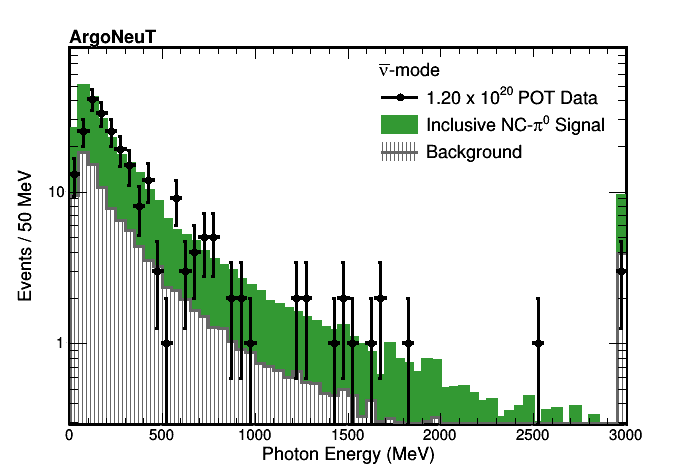}
    \includegraphics[scale=0.35]{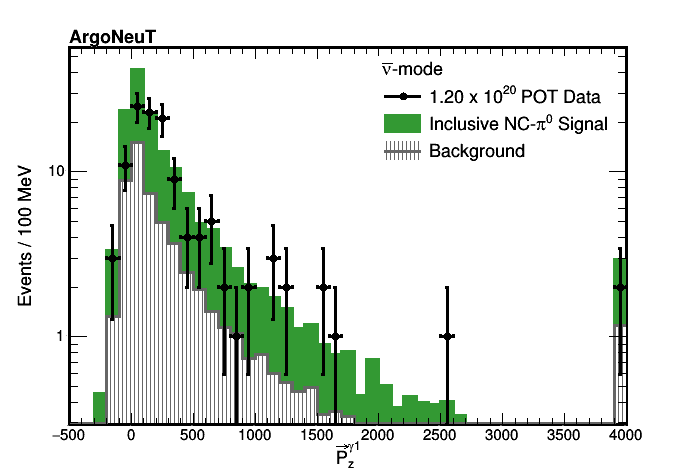}
    \includegraphics[scale=0.35]{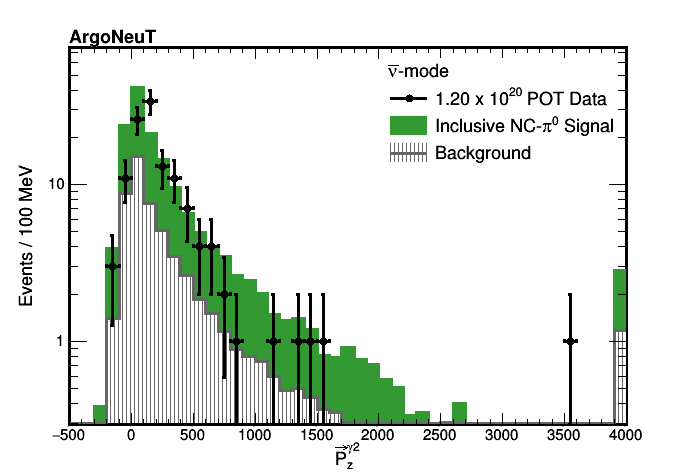}
    \includegraphics[scale=0.35]{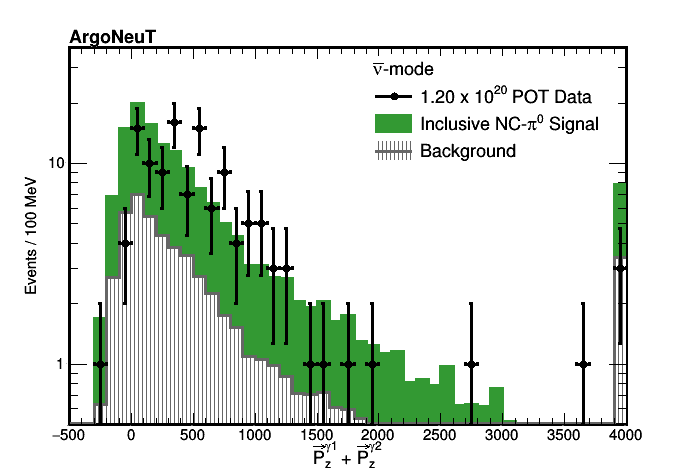}
    \caption{The photon energy for both photons in the event, the photon z momentum ($P_{z}$) for each photon ($\gamma_{1}, \gamma_{2}$, where the highest energy photon is listed first), and the sum of the z momentum for our data events with MC backgrounds scaled to 1.2$\times 10^{20}$ POT. These events include interactions from both $\nu$ and $\bar{\nu}$ scattering events. \label{fig:DataPhotonDistributions}}
 \end{center}
\end{figure}
\end{center}
\end{widetext}

The data presented in Fig. \ref{fig:PiDataPlots} and Fig. \ref{fig:DataPhotonDistributions} represent the observed number of events when compared to MC prediction as well as the kinematic distributions that go into these comparisons.

\section{Ratio of NC($\pi^{0}$) to CC}\label{sec:RatioNCtoCC}
Normalizing the NC$\pi^{0}$ production rate to the inclusive charged current rate measured by ArgoNeuT over the same running condition reduces many systematic uncertainties, particularly those associate with flux. This procedure also facilitates comparison with previous measurements of this ratio reported by other neutrino experiments \cite{SciBooNENCPi0, K2KNCPi0}.

In order to construct the ratio of NC$\pi^{0}$ to CC separately for neutrinos and anti-neutrinos, we divide the anti-neutrino beam into its components. For the charged current sample, \cite{CCInclusive}, the species of neutrino is determined by the MINOS-ND measuring the sign of the charged muon. In the case of NC$\pi^{0}$ production, no such data driven sign determination is possible. Instead we utilize MC to estimate the fraction of events coming from $\nu$ and $\bar{\nu}$ interactions. Our assumption that $f^{\nu}=74 \pm 15\%$ and $f^{\bar{\nu}}=26 \mp 15\%$ of NC$\pi^{0}$ originate from $\nu$ and $\bar{\nu}$ interactions, respectively, consistent with predictions from GENIE. GENIE also predicts that these fractions are independent of the $\nu$ and $\bar{\nu}$ energy. The uncertainty on this fraction is taken as a systematic and is described in Section \ref{sec:systematics}. This technique of separating the data sample utilizing the MC fraction of the beam means that the resulting fractions will be anti-correlated with one another through the systematic uncertainty of the beam content. 


Taking neutrino interactions as described in Eq. \ref{eqn:nuRatioOverview}, to illustrate the procedure, we define the numerator as:
\begin{equation} \label{eqn:NumeratorOfRatio}
N_{Events}(NC\pi^{0}) = \sum\limits_{i = bin} \frac{S_{i}^{\nu} - B_{i}^{\nu}}{\epsilon_{i}^{\nu}},
\end{equation}
where $S_{i}^{\nu}$ is the number of the signal events in a given bin from data that originates from neutrino ($\nu$) interactions and can be written as
\begin{equation}
S_{i}^{\nu} = f^{\nu} D_{i},
\end{equation}
where $f^{\nu}$ is the fraction of events coming from neutrino interactions (in this case approximately 74$\%$) and $D_{i}$ is the number of data events in that particular bin. In Eq. \ref{eqn:NumeratorOfRatio}, $B_{i}^{\nu}$ is the predicted background coming from $\nu$ interactions scaled to the appropriate protons on target (P.O.T). Finally, the term $\epsilon_{i}^{\nu}$ is the efficiency for neutrino induced NC$\pi^{0}$ events taken from MC is given by
\begin{equation}
\epsilon_{i}^{\nu} = \frac{\nu\mbox{ induced NC}\pi^{0}\mbox{'s passing all cuts in the i'th bin}}{\nu\mbox{ induced NC}\pi^{0}\mbox{'s generated in the i'th bin}}
\end{equation}
and is estimated to be $20.3\%$($15.4 \%$) for $\nu$ ($\bar{\nu}$) and is flat as a function of $P_{\pi^{0}}$ and $\cos(\pi^{0})$.   

An analogous procedure is followed for the anti-neutrino component of these interactions. Tab. \ref{tab:NCProductionSummary2} summarizes the results of the calculation from Eq. \ref{eqn:NumeratorOfRatio} for both the neutrino and anti-neutrino components of the anti-neutrino mode beam. The two sets of numbers for each type of interaction represent the parent distribution the data and background MC were drawn from (either $P_{\pi^{0}}$ (left hand side of \ref{fig:PiDataPlots}) or cosine of the $\pi^{0}$ (right hand side of Fig. \ref{fig:PiDataPlots})).

\begin{table}[htb]
	\begin{center}
	\resizebox{0.45\textwidth}{!}{%
	\begin{tabular}{|c|c|c|}
	\hline
	\multicolumn{3}{|c|}{\textbf{Efficiency Corrected NC$\pi^{0}$ Production}} \\
	\hline \hline
	 \textbf{Parent Distribution} & \textbf{Species} & \textbf{Events $\pm$ Stat. Error}    \\
	\hline
	$\pi^{0}$ Momentum & Neutrino & 311.4 $\pm$ 75.0 \\
	\hline
	$\pi^{0}$ Momentum & Anti-Neutrino & 97.5 $\pm$ 51.7 \\
	\hline
	Cosine $\pi^{0}$ & Neutrino & 328.5 $\pm$ 74.7 \\
	\hline
	Cosine $\pi^{0}$ & Anti-Neutrino & 104.2 $\pm$ 51.6 \\
	\hline
	\end{tabular}}
	\caption{Summary of the efficiency corrected semi-inclusive NC$\pi^{0}$ production on an argon target for the ArgoNeuT anti-neutrino data sample)} \label{tab:NCProductionSummary2}
	\end{center}
\end{table}

A similar procedure is followed for the charged current sample from Ref. \cite{CCInclusive} using Eq. \ref{eqn:NumeratorOfRatio} but this time for CC inclusive. Tab. \ref{tab:CCProductionSummary} summarizes the results of the calculation of the MC corrected charged current production for both the neutrino and anti-neutrino components of the anti-neutrino mode beam. The corrections applied to the CC inclusive sample include taking into account the various acceptances due to the neutrino and anti-neutrino components. The two sets of numbers for each type of interaction represent the parent distribution the data and background MC where drawn from (either the momentum of the lepton ($P_{\mu}$) or angle of the lepton ($\theta_{\mu}$)).

\begin{table}[htb]
	\begin{center}
	\resizebox{0.45\textwidth}{!}{%
	\begin{tabular}{|c|c|c|}
	\hline
	\multicolumn{3}{|c|}{\textbf{MC Corrected CC Production}} \\
	\hline \hline
	 \textbf{Parent Distribution} & \textbf{Species} & \textbf{Events $\pm$ Stat. Error}    \\
	\hline
	Lepton Momentum & Neutrino & 3425.9 $\pm$ 9.9 \\
	\hline
	Lepton Momentum & Anti-Neutrino & 2470.3 $\pm$ 4.4 \\
	\hline
	Lepton Angle & Neutrino & 3385.8 $\pm$ 14.4 \\
	\hline
	Lepton Angle & Anti-Neutrino & 2353.2 $\pm$ 9.3 \\
	\hline
	\end{tabular}}
	\caption{Summary of the MC corrected charged current production on an argon target for the ArgoNeuT anti-neutrino data sample)} \label{tab:CCProductionSummary}
	\end{center}
\end{table}

\subsection{Systematic Error}\label{sec:systematics}

\begin{itemize}
\item \textbf{MC estimation of the neutral current beam composition:}
As was stated earlier, we take from MC the fraction of the anti-neutrino beam that produces NC$\pi^{0}$ from neutrino or anti-neutrino interactions to be $f^{\nu}=74 \pm 15\%$ and $f^{\bar{\nu}}=26 \mp 15\%$. This error on the fraction is based on the difference in the fraction of the beam as estimated from the charged current inclusive sample and the fraction calculated from NC$\pi^{0}$ MC. Taking the numbers from Tab. \ref{tab:CCProductionSummary}, the fraction of the beam from $\nu$ and $\bar{\nu}$ is measured as
\begin{equation}
\frac{\nu\mbox{-CC}}{\mbox{Total CC}} = 59\% \mbox{ and } \frac{\bar{\nu}\mbox{-CC}}{\mbox{Total CC}} = 41\%
\end{equation}
Since there is no direct analogue to measure for NC$\pi^{0}$ production, the $\pm 15\%$ systematic covers this difference in the fraction calculated from the CC-inclusive data and those calculated from the NC$\pi^{0}$ MC. This conservatively assumes the maximum error in the beam composition is solely due to modelling of the beam.

\item \textbf{Ratio extraction from different parent histograms:}
The MC corrected ratio of NC$\pi^{0}$ to charged current production can be taken from either the momentum of the NC$\pi^{0}$ (CC-lepton) or from the cosine of the NC$\pi^{0}$ ($\theta$ of the CC-lepton). These two MC corrected histograms give slightly different results for the ratio, as summarized in Tab. \ref{tab:IndividualRatios}. The final answer is taken as the mean of the two results and the variation on the mean is taken as a systematic on the final ratio. 

\begin{table}[htb]
	\begin{center}
	\resizebox{0.44\textwidth}{!}{%
	\begin{tabular}{|c|c|c|}
	\hline
	\multicolumn{3}{|c|}{\textbf{NC$\pi^{0}$/CC Production Ratio}} \\
	\hline \hline
	 \textbf{Parent Distribution} & \textbf{Neutrino Ratio} & \textbf{Anti-Neutrino Ratio}    \\
	& $\pm$ (stat. only) & $\pm$ (stat. only) \\ 
	\hline
	Momentum & 0.091 $\pm$ 0.022 & 0.039 $\pm$ 0.021 \\
	\hline
	Angle &  0.097 $\pm$ 0.022 & 0.044 $\pm$ 0.022 \\
	\hline
	\end{tabular}}
	\caption{Summary of the ratios computed from either the momentum or angle distributions from Tab. \ref{tab:NCProductionSummary2} and Tab. \ref{tab:CCProductionSummary}} \label{tab:IndividualRatios}
	\end{center}
\end{table}

\item \textbf{Energy Correction Templates:}
The energy correction templates are allowed to vary between those derived for $\pi^{0} \rightarrow \gamma\gamma$ decays, shown in Fig. \ref{fig:ZTemplates}, Fig. \ref{fig:YTemplates}, and Fig. \ref{fig:XTemplates} in Appendix \ref{sec:EnergyCorrAppend} and the templates derived for single electrons. The variation in templates allows events to shift slightly bin-to-bin as well as in and out of the sample via the $M_{\gamma\gamma}$ cut defined in Section \ref{sec:RecoPi0Kin}. The variation in the ratio due to the differences in the energy templates is taken as a systematic on the final answer.
\end{itemize}

Tab. \ref{tab:SystematicErrors} summarizes the systematic errors and their relative magnitude on the final ratio as well as the total systematic taken on the computed ratio for neutrinos and anti-neutrinos.

\begin{table}[htb]
	\begin{center}
	\resizebox{0.45\textwidth}{!}{%
	\begin{tabular}{|c|c|c|}
	\hline
	\multicolumn{3}{|c|}{\textbf{Systematic Errors}} \\
	\hline \hline
	 \textbf{Source of the Error} & \textbf{ Error on } & \textbf{ Error on }    \\
	& \textbf{Neutrino Ratio} & \textbf{Anti-Neutrino Ratio}\\
	\hline
	Beam Composition & $\pm$ 0.014 & $\pm$ 0.006 \\
	\hline
	Parent Histogram & $\pm$ 0.006 & $\pm$ 0.005 \\
	\hline
	Energy Templates & $\pm$ 0.002 & $\pm$ 0.003 \\
	\hline
	\hline
	\textbf{Total Systematic Error} & \textbf{$\pm$ 0.015} & \textbf{$\pm$ 0.008} \\
	\hline
	\end{tabular}}
	\caption{Summary of the systematic errors on the ratio of NC production to CC production)} \label{tab:SystematicErrors}
	\end{center}
\end{table}

\subsection{Results}\label{sec:ratioresults}
Taking the mean of the two parent distributions and adding the systematics the final ratios for both neutrino and anti-neutrino interactions are
\begin{equation}
\frac{\sigma_{\nu}(NC\pi^{0})}{\sigma_{\nu}(CC)} = 0.094 \pm 0.022\mbox{(stat.)} \pm 0.015 \mbox{(sys.)}
\end{equation}
and
\begin{equation}
\frac{\sigma_{\bar{\nu}}(NC\pi^{0})}{\sigma_{\bar{\nu}}(CC)} = 0.042 \pm 0.022\mbox{(stat.)} \pm 0.008 \mbox{(sys)}
\end{equation}
for neutrinos with a mean energy of 9.6 GeV and anti-neutrinos with a mean energy of 3.6 GeV. The total inclusive ratio calculated from the sum of the neutrino and anti-neutrino component is 
\begin{equation}
\frac{\sigma(NC\pi^{0})}{\sigma(CC)} = 0.136 \pm 0.031\mbox{(stat.)} \pm 0.017 \mbox{(sys)}.
\end{equation}

The result is plotted on Fig. \ref{fig:RatioPlot} showing the computed ratio of NC$\pi^{0}$ production to inclusive CC scattering as computed using GENIE and NuWro neutrino simulations on an argon target. To compute the ratio from GENIE and NuWro we take the total neutral current production cross-section and scale it by the fraction of those events that produce $\geq 1 \pi^{0}$ in the event and divide by the charged current inclusive cross-section.

\begin{widetext}
\begin{center}
\begin{figure}[htb]
  \begin{center}
    \includegraphics[scale=0.30]{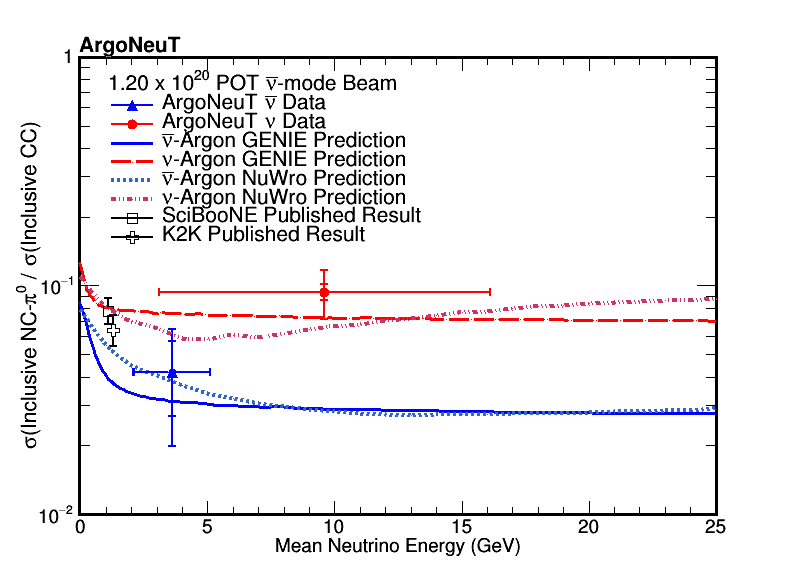}
    \includegraphics[scale=0.30]{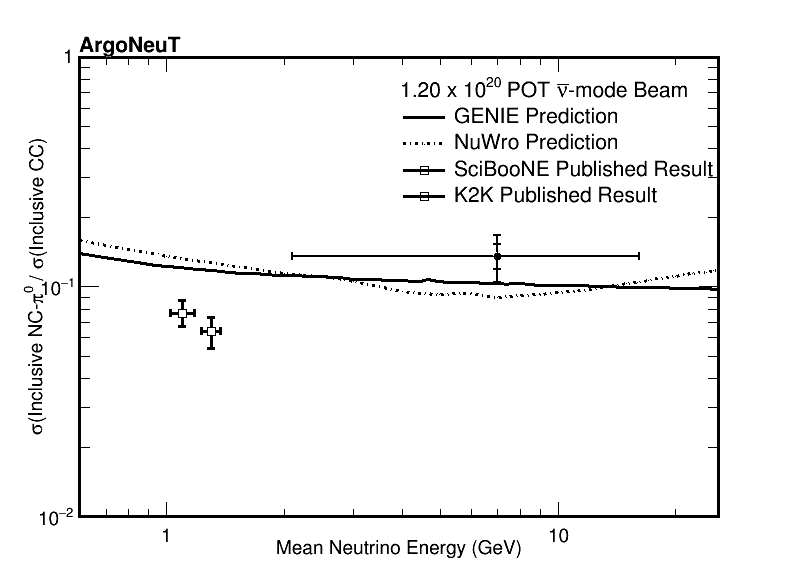}
    \caption{Ratio of the NC$\pi^{0}$ production to inclusive CC scattering cross-sections for both neutrino (red) and anti-neutrino (blue) scattering as measured by ArgoNeuT and as computed using the GENIE and NuWro neutrino simulation on an argon target. For reference, the results obtained by the SciBooNE collaboration for a neutrino beam with a mean energy of 1.1 GeV on a polystyrene target (C8H8) as well as the results from the K2K collaboration for a neutrino beam with a mean energy of 1.3 GeV on a water target are shown. The ArgoNeuT measurement is shown with statistical and total error bars. \label{fig:RatioPlot}}
 \end{center}
\end{figure}
\end{center}
\end{widetext}

Comparing the ArgoNeuT measured values for the ratio of NC$\pi^{0}$/CC to those reported by the SciBooNE collaboration for a neutrino beam with a mean energy of 1.1 GeV on a polystyrene target (C8H8) (0.077 $\pm 0.5$(stat.) $\pm 0.5$ (sys.)) \cite{SciBooNENCPi0} and with the K2K collaboration for a neutrino beam with a mean energy of 1.3 GeV on a water target (0.064 $\pm$ 0.001 (stat.) $\pm$ 0.007 (sys.))\cite{K2KNCPi0} we find that the ArgoNeuT ratio is slightly higher for the higher energy beam.

\section{Flux Averaged NC$\pi^{0}$ Cross-Section}\label{sec:FluxAvgXSection}
In addition to calculating the ratio of NC$\pi^{0}$ to CC, one can also calculate the flux averaged absolute cross-section for the NC$\pi^{0}$ production. This is defined as
\begin{equation} \label{eqn:NuCrossSection}
\sigma_{\nu}(NC\pi^{0}) = \sum\limits_{i = bin} \frac{S_{i}^{\nu} - B_{i}^{\nu}}{\epsilon_{i}^{\nu}\Phi_{\nu}N_{Targets}}
\end{equation}
and 
\begin{equation}
\sigma_{\bar{\nu}}(NC\pi^{0}) = \sum\limits_{i = bin} \frac{S_{i}^{\bar{\nu}} - B_{i}^{\bar{\nu}}}{\epsilon_{i}^{\bar{\nu}}\Phi_{\bar{\nu}}N_{Targets}},
\end{equation}
where $S_{i}^{\nu / \bar{\nu}}$, $B_{i}^{\nu / \bar{\nu}}$ and $\epsilon_{i}^{\nu / \bar{\nu}}$ are defined just as before and $N_{Targets}$ represents the number of argon nuclei in the fiducial volume and $\Phi_{\nu / \bar{\nu}}$ is the neutrino/anti-neutrino flux exposure given in Tab. \ref{tab:Fluxes}.

Similar to Section \ref{sec:RatioNCtoCC}, the component of the sample coming from $\nu$ and $\bar{\nu}$ is derived using the MC. The total integrated flux for the $\nu$ and $\bar{\nu}$ components of the anti-neutrino beam is taken from Ref. \cite{CCInclusive}.
\subsection{Cross-Section Systematic Error}\label{sec:CXsystematics}
In addition to the systematics described before, three new systematics are present when the result is interpreted as an integrated cross-section. The integrated flux is assigned a flat 11$\%$ uncertainty that accounts for the uncertainty in hadron production and beam line modelling and has been used in previous ArgoNeuT analyses \cite{CCInclusive, Coherent}. Uncertainties in the number of targets and P.O.T. are taken into account as well, although these contribute only at the few percent level. The full description of the systematics applied to the integrated cross-section are given in Tab. \ref{tab:SystematicErrorsXSec}.
\newline
\newline

\begin{table}[htb]
	\begin{center}
	\resizebox{0.45\textwidth}{!}{%
	\begin{tabular}{|c|c|c|}
	\hline
	\multicolumn{3}{|c|}{\textbf{Systematic Errors}} \\
	\hline \hline
	 \textbf{Source of the Error} & \textbf{$\%$ Error on $\sigma(\nu)$} & \textbf{$\%$ Error on $\sigma(\bar{\nu})$ }    \\
	\hline
	Beam Composition & $\pm$ 15$\%$ & $\pm$ 15$\%$ \\
	\hline
	Flux Normalization & $\pm$ 11$\%$ & $\pm$ 11$\%$ \\
	\hline
	Parent Histogram & $\pm$ 4.3$\%$ & $\pm$ 8.1$\%$ \\
	\hline
	Number of Targets & $\pm$ 2$\%$ & $\pm$ 2$\%$ \\
	\hline
	Energy Templates & $\pm$ 1$\%$ & $\pm$ 1$\%$ \\
	\hline
	P.O.T. & $\pm$ 1$\%$ & $\pm$ 1$\%$ \\
	\hline
	\hline
	\textbf{Total Systematic Error} & \textbf{$\pm$ 18.7$\%$} & \textbf{$\pm$ 20.4$\%$} \\
	\hline
	\end{tabular}}
	\caption{Summary of the systematic errors on the ratio of NC production to CC production)} \label{tab:SystematicErrorsXSec}
	\end{center}
\end{table}

\subsection{Cross-Section Results}\label{sec:CXresults}
Taking the mean of the two parent distributions and adding the full systematics, the cross-section for both neutrino and anti-neutrino interactions is measured to be
\begin{equation}
\sigma_{\nu}(NC\pi^{0}) = (7.1 \pm 1.7\mbox{(stat.)} \pm 1.3 \mbox{(sys.)})\times10^{-40}\mbox{cm}^{2}
\end{equation}
and
\begin{equation}
\sigma_{\bar{\nu}}(NC\pi^{0}) = (0.5 \pm 0.2\mbox{(stat.)} \pm 0.1\mbox{(sys.)})\times10^{-40}\mbox{cm}^{2}
\end{equation}
per argon nucleon with anti-neutrinos at a mean energy of 3.6 GeV and neutrinos at a mean energy of 9.6 GeV. These results are plotted in Fig. \ref{fig:CrossSectionPlot} as well as the inclusive result taken by adding together the contributions due to the two components. 
\begin{equation}
\sigma(NC\pi^{0}) = (7.6 \pm 1.7\mbox{(stat.)} \pm 1.4\mbox{(sys.)})\times10^{-40}\mbox{cm}^{2}
\end{equation}
These results are shown with a comparison to the predictions from the GENIE and NuWro event generators. The predictions of GENIE and NuWro are consistent with the ArgoNeuT measurements.

\begin{widetext}
\begin{center}
\begin{figure}[htb]
  \begin{center}
    \includegraphics[scale=0.30]{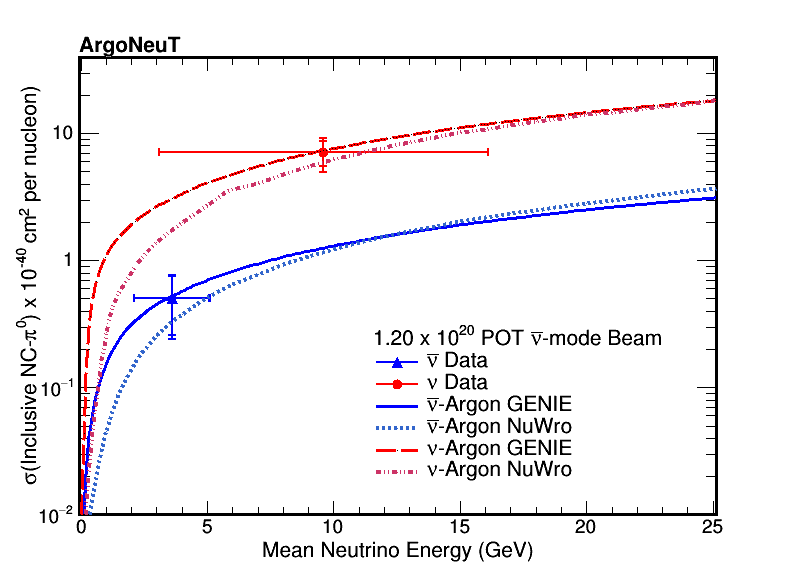}
    \includegraphics[scale=0.30]{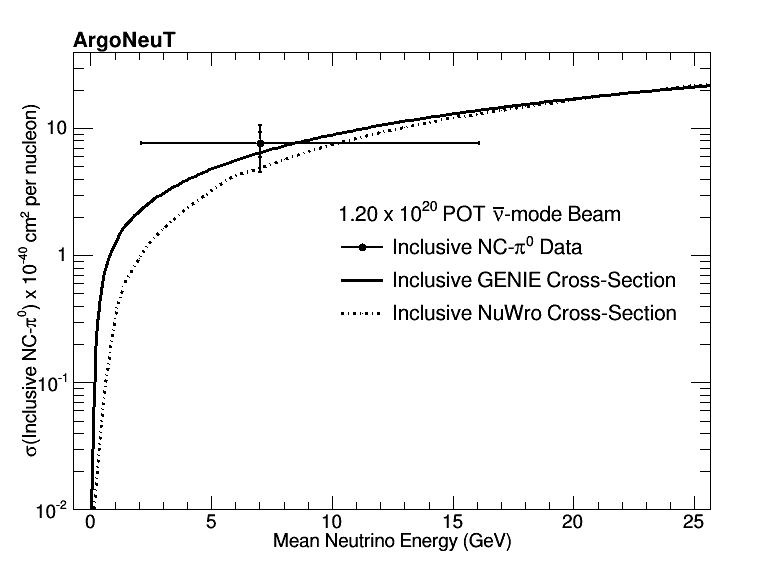}
    \caption{The NC$\pi^{0}$ production cross-section for both neutrino (red) and anti-neutrino (blue) scattering on argon as measured in ArgoNeuT and as predicted by the GENIE (solid) and NuWro (dashed) generators. The ArgoNeuT measurement has been flux averaged and is shown with both statistical and total uncertainties. \label{fig:CrossSectionPlot}}
 \end{center}
\end{figure}
\end{center}
\end{widetext}

Since this is the first time this process has been measured on an argon nuclei, comparison with previous NC$\pi^{0}$ data proves difficult. For example, MiniBooNE (\cite{MiniBooNENCPi02}) reports an absolute cross-section for single $\pi^{0}$ neutral current production of (4.76$\pm$0.05 (stat)$\pm$ 0.76 (sys))$\times$10$^{-40}$ cm$^2$/nucleon at a mean energy for neutrinos of 808 MeV and (1.48$\pm$0.05 (stat)$\pm$0.23 (sys))$\times$10$^{-40}$ cm$^2$/nucleon at a mean energy for anti-neutrinos of 664 MeV. While both these results are of the same order as the ArgoNeuT reported cross-section, the MiniBooNE's target nuclei was CH$_{2}$, and the scaling to the more dense nuclei of argon is not well understood.

\section{Discussion}
In order to perform this measurement many novel techniques for identifying and reconstructing electromagnetic showers in a small volume LArTPC were employed and are presented here along side the measurement of neutral current $\pi^{0}$ production. One such technique presented here is a method to account for the missing energy of the photon shower when it is loss outside the fiducial boundary. By utilizing the well known electromagnetic shower development in liquid argon with the visible portion of the shower in the detector it is possible to correct back the loss energy. Improvements on this technique will enable future larger LArTPC experiments, such as MicroBooNE, SBND, and ICARUS, to increase their acceptance of NC$\pi^{0}$ production within their detectors. 

The $\pi^{0}$ kinematic distributions presented in Section \ref{sec:RecoPi0Kin} agree within the experimental uncertainties in both rate the shape with the predicted simulations. Future larger LArTPC's will have the capability to improve greatly on these measurements and probe with even better resolution these distributions.

The interpretation of these kinematic distributions as the ratio of the efficiency corrected neutral current $\pi^{0}$ production to total inclusive charged current cross-section on argon as well as flux averaged absolute cross-section are the first of their kind done on an argon target. Both the ratio and the flux averaged absolute cross-section are found to be consistent with predictions from simulation as well as previous data measurements. The ArgoNeuT measurements provide information on the $A$ dependence of neutrino cross sections that may prove helpful when attempting to estimate future cross-section uncertainties for LArTPC's.

\section{Conclusions}
In conclusion, the ArgoNeuT Collaboration reports the first measurement of neutrino and anti-neutrino semi-inclusive neutral current $\pi^{0}$-production on an argon target. We present: 1) kinematic distribution of the $\pi^{0}$ mesons produced in the neutrino-argon interaction, 2) the ratio of the GENIE and NuWro event generators corrected neutral current $\pi^{0}$ production to the total inclusive charged current production and 3) the flux averaged cross-section for neutrinos with a mean energy of 9.6~GeV and for anti-neutrinos with a mean energy of 3.6~GeV. Both the ratio and the cross-section are broken into a contribution from the neutrino and the anti-neutrino profile based on a Monte Carlo estimate of the beam composition. Both of the values obtained are consistent with predictions from the GENIE and NuWro neutrino generator Monte Carlo. 

\acknowledgments

ArgoNeuT gratefully acknowledges the cooperation of the MINOS collaboration in providing data for the use in this analysis. We would also like to acknowledge the support of Fermilab (Operated by Fermi Research Alliance, LLC under Contract No. De‐AC02‐07CH11359 with the United States Department of Energy), the Department of Energy, and the National Science Foundation in the construction, operation and data analysis of ArgoNeuT. 
\appendix

\section{Electromagnetic Energy Corrections inside the ArgoNeuT TPC} \label{sec:EnergyCorrAppend}
In this appendix we provide a detailed overview of the procedures used to generate the energy correction templates used in this analysis and initially described in Section \ref{sec:EnergyCorr}. We begin with a discussion of the angle reconstruction seen for single particle $\pi^{0}$ MC that becomes the basis of the template approach to applying energy corrections. This is followed by a detailing of the various energy correction procedures and templates that are applied. Finally, a summary of the cross-checks that were performed showing the robustness of the energy corrections is presented.

\subsection{Momentum Hypothesis from reconstructed angles}\label{sec:AngleAppend}
Utilizing one of the great strength's of the LArTPC technology, its fine grain tracking information, allows for reconstruction of the angle of an electromagnetic shower despite very little of the shower being contained. This was demonstrated in Section \ref{sec:MCObservables} and Fig. \ref{fig:AngleDepo3} shows the performance of the shower reconstruction on simulated $\pi^{0}\rightarrow\gamma\gamma$ events using the hand-scan and automated shower reconstruction tools described in Section \ref{sec:EventSelection}. These tools faithfully reconstruct the true angle of the photon shower and thus the angular reconstruction can be trusted to represent the underlying true photon angles. This information becomes very useful when attempting to correct back the energy loss of a shower due to poor containment. Specifically, it provides another method to calculate a hypothesis for the initial momentum the $\pi^{0}$ based on the opening angle between the pair of photons in the event ($\theta_{\gamma\gamma}$). 

\begin{widetext}
\begin{center}
\begin{figure}[htb]
  \begin{center}
    \includegraphics[scale=0.45]{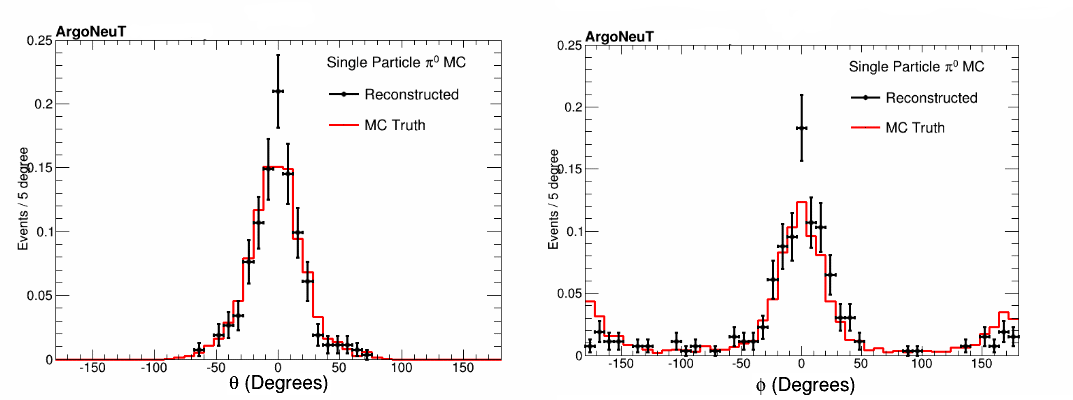}
    \caption{Single particle Monte Carlo angles that have been fully reconstructed using the shower hand-scan + automated shower tools. \label{fig:AngleDepo3}}
 \end{center}
\end{figure}
\end{center}
\end{widetext}

\begin{figure}[htb]
  \begin{center}
    \includegraphics[width=0.48\textwidth]{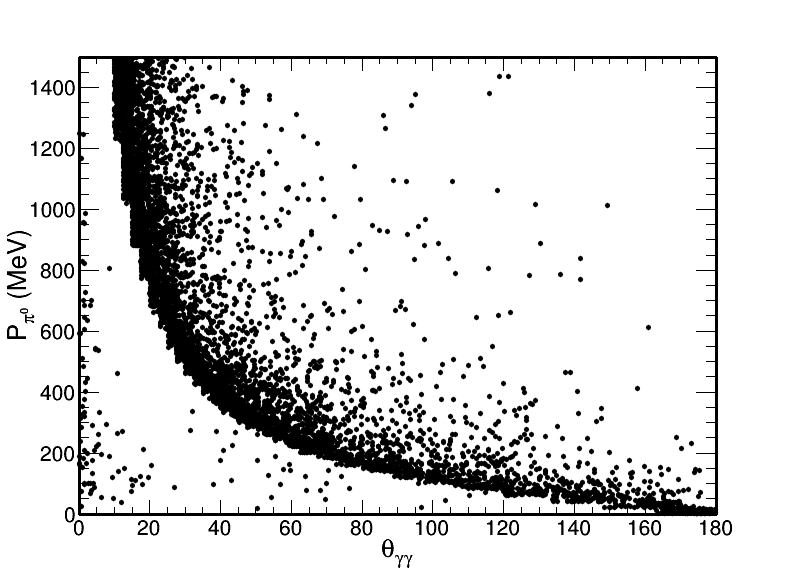}
    \caption{Momentum versus opening angle shown as a 2-d distribution. The profile fit with a function shown in Fig. \ref{fig:MomentumVsAngle} that allows a hypothesis for the upper bound on the momentum of the $\pi^{0}$ system. \label{fig:MomentumVsAngle2d}}
 \end{center}
\end{figure}

As mentioned previously, Fig. \ref{fig:MomentumVsAngle} shows the strong correlation between the initial momentum of the $\pi^{0}$ and the opening angle $\theta_{\gamma\gamma}$.  Fig. \ref{fig:MomentumVsAngle2d} provides the two dimensional (2D) distribution from which the fitted function, Eq. \ref{eqn:AngleMomentum}, that defines the $\pi^{0}$ momentum as a function of the opening angle between the photons. The constants from the fit are $C_{0}$ = 2202.3, $C_{1}$ = -94.9, $C_{2}$ = 2.1, $C_{3}$ = -0.025, $C_{4}$ = 0.00017, $C_{5}$ = -6.0$\times 10^{-7}$, $C_{6}$ = 8.5$\times 10^{-10}$.

Comparing the difference between the momentum of the $\pi^{0}$ calculated utilizing Eq. \ref{eqn:AngleMomentum} and the momentum utilizing Eq. \ref{eqn:EnergyMomentum}, as is done in Figure \ref{fig:MomentumVsAngle}, shows that the angle method reconstructs the momentum to within $\pm$30$\%$ with a slight bias towards overestimating the momentum. While this accuracy is insufficient to directly extract the physics of the $\pi^{0}$ system, using this as a starting point for our energy correction calculation is sufficient. By being able to bound the hypothesis for the momentum of the $\pi^{0}\rightarrow\gamma\gamma$ we can then go about deriving energy corrections using topological information.

\begin{figure}[htb]
  \begin{center}
    \includegraphics[width=0.48\textwidth]{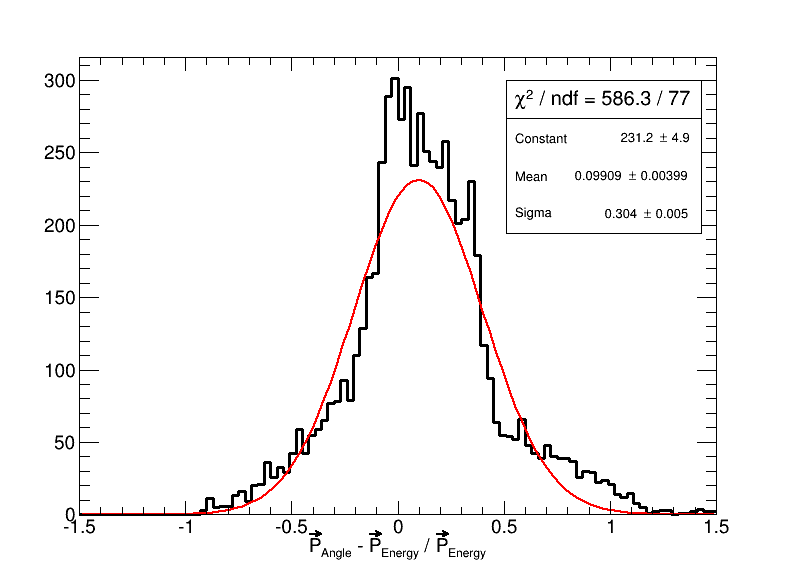}
    \caption{Difference between $\pi^{0}$  momentum estimated from $\gamma\gamma$ opening angle and true $\pi^{0}$ momentum, from MC. This method defined in Eq. \ref{eqn:AngleMomentum} reconstructs the momentum of the $\pi^{0}$ system to within $\pm$30$\%$. \label{fig:MomentumTrue_vs_MomentumAngle}}
 \end{center}
\end{figure}


\subsection{Linear Correction}\label{sec:LinearCorrAppend}
\begin{figure}[htb]
  \begin{center}
    \includegraphics[width=0.48\textwidth]{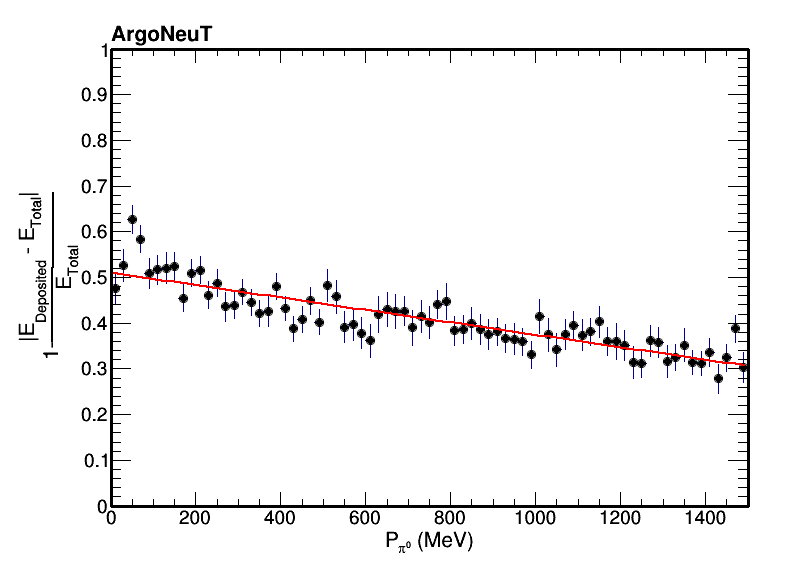}
    \caption{Template for the linear energy correction applied to the photons based on the $P_{\pi^{0}}$. \label{fig:FlatCorrection}}
 \end{center}
\end{figure}

Fig. \ref{fig:FlatCorrection} shows the fraction of the contained energy (as defined in Eq. \ref{eqn:EnergyDep}) as a function of the initial true momentum of the $\pi^{0}$.

Utilizing the initial momentum hypothesis formed using Eq. \ref{eqn:AngleMomentum} for the $\pi^{0}$ ($P_{\pi^{0}}^{I}$) the linear correction that is attempted to be applied to the photons has the form
\begin{equation}
E^{Flat Corr}_{\gamma^{i}} = (P_{\pi^{0}}^{I} \times C_{0}^{Linear}) + C_{1}^{Linear},
\end{equation}
where the constants $C_{i}$ are obtained from the fit of Fig. \ref{fig:FlatCorrection} and found to be $C_{0} = 0.51$ and $C_{1} = $-1.4e$\times 10^{-4}$. 

Thus the first correction that is tried is a simple linear correction applied to the photon energy as defined in Eq. \ref{eqn:FlatCorr}.
\begin{equation}\label{eqn:FlatCorr}
E_{\gamma^{i}} = E^{0}_{\gamma^{i}} + E^{0}_{\gamma^{i}} \times E^{Flat Corr}_{\gamma^{i}},
\end{equation}
where $E^{0}_{\gamma^{i}}$ is the original energy of the photon (uncorrected). If this correction for any of the photon pairs does not cause the momentum of the $\pi^{0}$ system as calculated using Eq. \ref{eqn:EnergyMomentum} to exceed the hypothesis for the momentum as calculated using Eq. \ref{eqn:AngleMomentum}, then the correction is kept and stored. Otherwise the correction becomes
\begin{equation}
E^{Flat Corr}_{\gamma^{i}} = 0.0.
\end{equation}

\subsection{X, Y, Z Topology Corrections}\label{sec:XYZAppend}
\begin{figure}[htb]
  \begin{center}
    \includegraphics[width=0.48\textwidth]{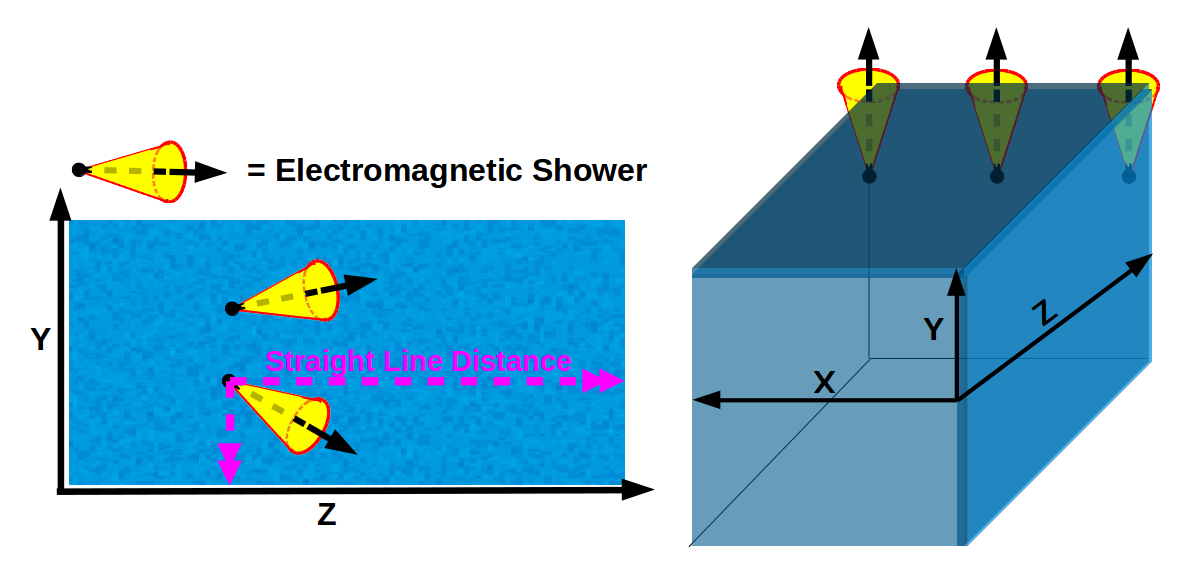}
    \caption{Illustration of the topological layout of an electromagnetic shower (represented as a yellow cone) and the direction the shower is propagating. The left hand side represents two showers created at the same Z location but with different Y initial positions. The right hand side shows three showers all created with the same Y-Z location, but with different initial X positions. Each of these cases demonstrates the topology of the shower matters when calculating the shower containment. This analysis uses the ``straight line'' distance to the wall in X, Y, and Z independently to calculate an energy correction.\label{fig:DistanceCalc}}
 \end{center}
\end{figure}

The next set of corrections that are applied are based on where the shower vertex is found inside the detector. The amount of energy that is deposited inside the detector is strongly correlated with where the photon first converts in X,Y,Z. Fig. \ref{fig:DistanceCalc} is a schematic demonstrating how the geometry and topological distribution of the shower, where the shower is created and what direction the shower is pointing, will determine how much of the shower is contained. For example, as is illustrated on the left of Fig. \ref{fig:DistanceCalc}, a shower that is created at the front (small Z) of the detector but is pointed down (negative Y) will have worse shower containment when compared to a shower that is created at the same point but is pointed up (positive Y). 

Moreover, as is illustrated on the right side of Fig. \ref{fig:DistanceCalc}, three different showers all created at the same Z location and pointing upward ($\phi\sim 90$) can have vastly different containment based on where they are in X (drift direction). For this reason, when we calculate the distance to the nearest wall in X, Y, and Z we use the ``straight line'' distance (illustrated as dashed line in Fig. \ref{fig:DistanceCalc}) instead of the ``pointing line'' (illustrated with the arrow and referred to at the $R$ distance). By taking into account all three spatial variables separately and in turn there is a greater chance of correcting the component of the energy loss that is due to the spatial variable that matters, instead of rolling all the information into one variable ($R$).

Taking the origin of the detector to be on the beam right, center of the upstream face and defining the shower vertex relative to this position we are able to calculate the distance of the ``closest'' boundary in $X$, $Y$, and $Z$ using the following relations:
\begin{center}
$Z_{Boundary} = 90$ if $-90 \leq \phi \leq 90$

$Z_{Boundary} = 0$ if $90 \leq \phi \leq 180$ or $-180 \leq \phi \leq -90$

$Y_{Boundary} = 20$ if $\phi \geq 0$

$Y_{Boundary} = -20$ if $\phi \leq 0$

$X_{Boundary} = 0.0$ if $\theta \geq 0$

$X_{Boundary} = 50.0$ if $\theta \leq 0$
\end{center}

To correct back the energy loss due to the topological location of the electromagnetic shower, we plot the fraction of energy contained (as defined in Eq. \ref{eqn:EnergyDep}) versus the distance to the X,Y, and Z boundary. We then fit the projection of these distributions with a polynomial that minimizes the $\chi^{2}$/NDF to give us a functional form for the energy correction. For each photon we calculate its distance to any boundary and apply back an energy correction based on this function of $x$, $y$, and $z$ distance.

\begin{center}
\textbf{Z Distance to a Boundary Template}
\end{center}

\begin{figure}[h]
  \begin{center}
    \includegraphics[width=0.48\textwidth]{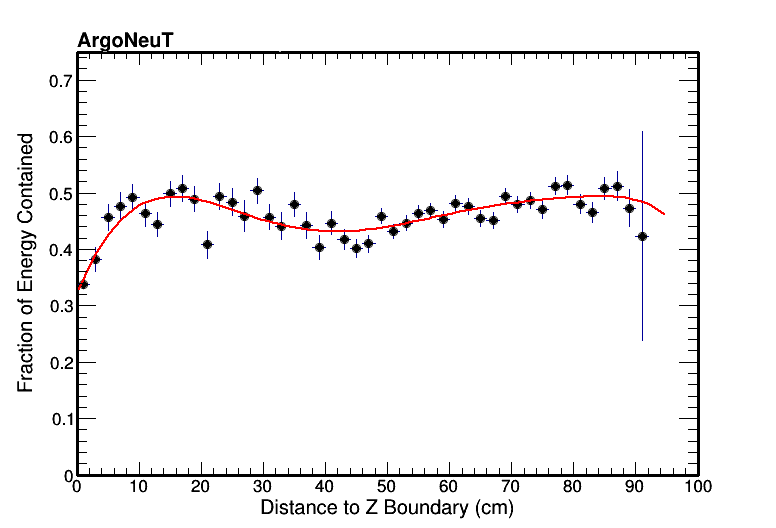}
    \caption{Templates of the fraction of the energy contained as a function of the shower's distance Z boundarywith the constants in the polynomial fit given by $C_{0}^{z}$ = 0.32,  $C^{z}_{1}$ =  0.028, $C^{z}_{2}$ = -0.0014, $C^{z}_{3}$ = 2.5e-05, $C^{z}_{4}$ = -2.8e-08, $C^{z}_{5}$ = -2.4e-09, $C^{z}_{6}$ = -1.6e-12, $C^{z}_{7}$ = 3.3e-13, $C_{z}^{8}$ = -1.8e-15 \label{fig:ZTemplates}}
 \end{center}
\end{figure}

Fig. \ref{fig:ZTemplates} shows the polynomial fit to the plot of the average fraction of the energy contained as a function of the distance to the Z boundary. The best fit returns an eighth degree polynomial based on the distance of the shower vertex to the nearest Z boundary defined by the Equation below

\begin{equation}
E^{Z Dist Corr}_{\gamma^{i}} = \sum_{i=0}^{8} C_{i}^{z} Z_{Dist}^{i},
\end{equation}
where the constants are read from the polynomial fit for the energy of a given photon becomes

\begin{equation}
E_{\gamma^{i}} = E^{0}_{\gamma^{i}} + E^{Flat Corr}_{\gamma^{i}} + (E^{0}_{\gamma^{i}} \times E^{Z Dist Corr}_{\gamma^{i}}),
\end{equation}
where $E^{0}_{\gamma^{i}}$ is the original energy of the photon (uncorrected) and $E^{Flat Corr}_{\gamma^{i}}$ is defined in Eq. \ref{eqn:FlatCorr}. If this correction for any of the photon pairs does not cause the momentum of the $\pi^{0}$ system as calculated using Eq. \ref{eqn:EnergyMomentum} to exceed the hypothesis for the momentum as calculated using Eq. \ref{eqn:AngleMomentum}, then the correction is kept and stored as
\begin{equation}\label{eqn:ZCorr}
E^{Z Corr}_{\gamma^{i}} = E^{0}_{\gamma^{i}} \times E^{Z Dist Corr}_{\gamma^{i}}.
\end{equation}
If it does exceed the momentum hypothesis then it is stored as
\begin{equation}
E^{Z Corr}_{\gamma^{i}} = 0.0.
\end{equation}

\begin{center}
\textbf{Y Distance to a Boundary Template}
\end{center}

\begin{figure}[h]
  \begin{center}
    \includegraphics[width=0.48\textwidth]{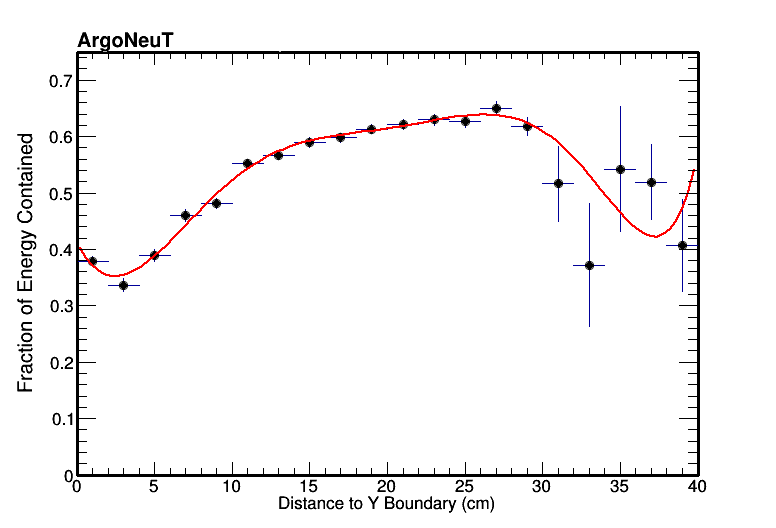}
    \caption{Templates of the fraction of the energy contained as a function of the shower's distance Y boundary with the constants in the polynomial fit given by $C_{0}^{y}$ = 0.41, $C_{1}^{y}$ = -0.054, $C_{2}^{y}$ = 0.014, $C_{3}^{y}$ = -0.0010, $C_{4}^{y}$ = 2.2$\times 10^{-5}$, $C_{5}^{y}$ = 6.3$\times 10^{-7}$, $C_{6}^{y}$ = -3.3$\times 10^{-8}$, and $C_{7}^{y}$ = 3.7$\times 10^{-10}$ \label{fig:YTemplates}}
 \end{center}
\end{figure}

Fig. \ref{fig:YTemplates} shows the polynomial fit to the plot of the average fraction of the energy contained as a function of the distance to the Y boundary. The best fit returns a seventh degree polynomial based on the distance of the shower vertex to the nearest Y boundary defined by the Equation below

\begin{equation}
E^{Y Dist Corr}_{\gamma^{i}} = \sum_{i=0}^{7} C_{i}^{y} Y_{Dist}^{i},
\end{equation}
where the constants are read from the polynomial fit. Now the calculation for the energy of a given photon becomes

\begin{equation}
E_{\gamma^{i}} = E^{0}_{\gamma^{i}} + E^{Flat Corr}_{\gamma^{i}} + E^{Z Corr}_{\gamma^{i}} + (E^{0}_{\gamma^{i}} \times E^{Y Dist Corr}_{\gamma^{i}}),
\end{equation}
where $E^{0}_{\gamma^{i}}$ is the original energy of the photon (uncorrected) and $E^{Flat Corr}_{\gamma^{i}}$ is defined in Eq. \ref{eqn:FlatCorr} and $E^{Z Corr}_{\gamma^{i}}$ is defined in Eq. \ref{eqn:ZCorr}. If this correction for any of the photon pairs does not cause the momentum of the $\pi^{0}$ system as calculated using Eq. \ref{eqn:EnergyMomentum} to exceed the hypothesis for the momentum as calculated using Eq. \ref{eqn:AngleMomentum}, then the correction is kept and stored as

\begin{equation}\label{eqn:YCorr}
E^{Y Corr}_{\gamma^{i}} = E^{0}_{\gamma^{i}} \times E^{Y Dist Corr}_{\gamma^{i}}.
\end{equation}

If it does exceed the momentum hypothesis then it is stored as:

\begin{equation}
E^{Y Corr}_{\gamma^{i}} = 0.0.
\end{equation}

\begin{center}
\textbf{X Distance to a Boundary Template}
\end{center}

\begin{figure}[h]
  \begin{center}
    \includegraphics[width=0.48\textwidth]{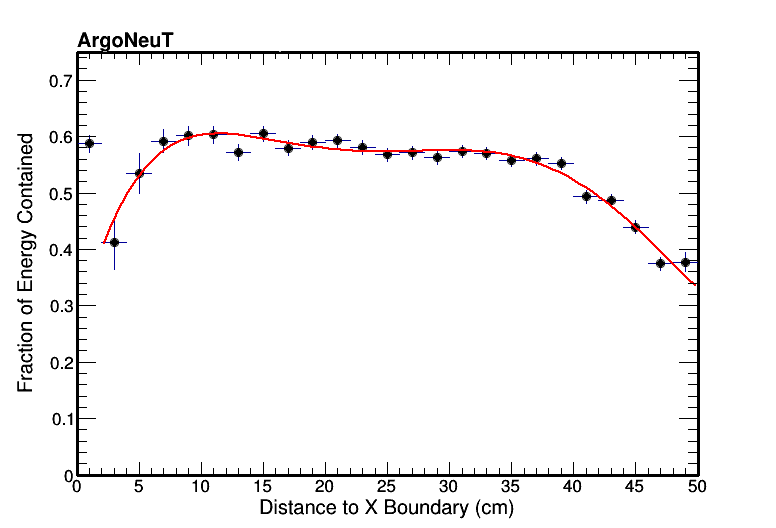}
    \caption{Templates of the fraction of the energy contained as a function of the shower's distance X boundary with the constants in the polynomial fit given by  $C_{0}^{x}$ = 0.25 , $C_{1}^{x}$~=~0.088, $C_{2}^{x}$~=~-0.0078, $C_{3}^{x}$~=~0.00031, $C_{4}^{x}$~=~-5.6$\times 10^{-6}$, and $C_{5}^{x}$~=~3.7$\times 10^{-8}$\label{fig:XTemplates}}
 \end{center}
\end{figure}

Fig. \ref{fig:XTemplates} shows the polynomial fit to the plot of the average fraction of the energy contained as a function of the distance to the X boundary. The best fit returns a fifth degree polynomial based on the distance of the shower vertex to the nearest X boundary defined by the Equation below

\begin{equation}
E^{X Dist Corr}_{\gamma^{i}} = \sum_{i=0}^{5} C_{i}^{x} X_{Dist}^{i},
\end{equation}
where the constants are read from the polynomial fit. Now the calculation for the energy of a given photon becomes

\begin{eqnarray*}
E_{\gamma^{i}} & =  & E^{0}_{\gamma^{i}} + E^{Flat Corr}_{\gamma^{i}} + E^{Z Corr}_{\gamma^{i}} \\
& & + E^{Y Corr}_{\gamma^{i}} + (E^{0}_{\gamma^{i}} \times E^{X Dist Corr}_{\gamma^{i}}),
\end{eqnarray*}
where $E^{0}_{\gamma^{i}}$ is the original energy of the photon (uncorrected) and $E^{Flat Corr}_{\gamma^{i}}$ is defined in Eq. \ref{eqn:FlatCorr} and $E^{Z Corr}_{\gamma^{i}}$ is defined in Eq. \ref{eqn:ZCorr} and $E^{Y Corr}_{\gamma^{i}}$ is defined in Eq. \ref{eqn:YCorr}. If this correction for any of the photon pairs does not cause the momentum of the $\pi^{0}$ system as calculated using Eq. \ref{eqn:EnergyMomentum} to exceed the hypothesis for the momentum as calculated using Eq. \ref{eqn:AngleMomentum}, then the correction is kept and stored as

\begin{equation}\label{eqn:XCorr}
E^{X Corr}_{\gamma^{i}} = E^{0}_{\gamma^{i}} \times E^{X Dist Corr}_{\gamma^{i}}.
\end{equation}

If it does exceed the momentum hypothesis than it is stored as

\begin{equation}
E^{X Corr}_{\gamma^{i}} = 0.0.
\end{equation}

Then end result of all these corrections is for any single photon to have its energy calculated as:

\begin{equation}\label{eqn:finalPhotonE}
E_{\gamma^{i}} = E^{0}_{\gamma^{i}} + E^{Flat Corr}_{\gamma^{i}} + E^{Z Corr}_{\gamma^{i}} + E^{Y Corr}_{\gamma^{i}} + E^{X Corr}_{\gamma^{i}}.
\end{equation}

The final results of these corrections appear to be insensitive the order they are applied as long as the constraint based on the $\pi^{0}$ momentum is applied. In this analysis, the corrections are applied as they are laid out here.

\subsection{Energy Correction Results}\label{sec:ECorrResultsAppend}
\begin{figure}[htb]
  \begin{center}
    \includegraphics[width=0.48\textwidth]{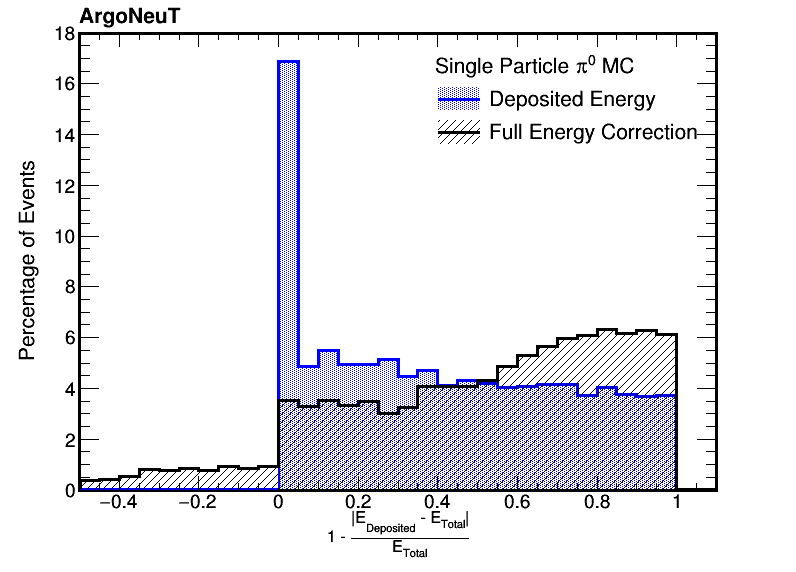}
    \includegraphics[width=0.48\textwidth]{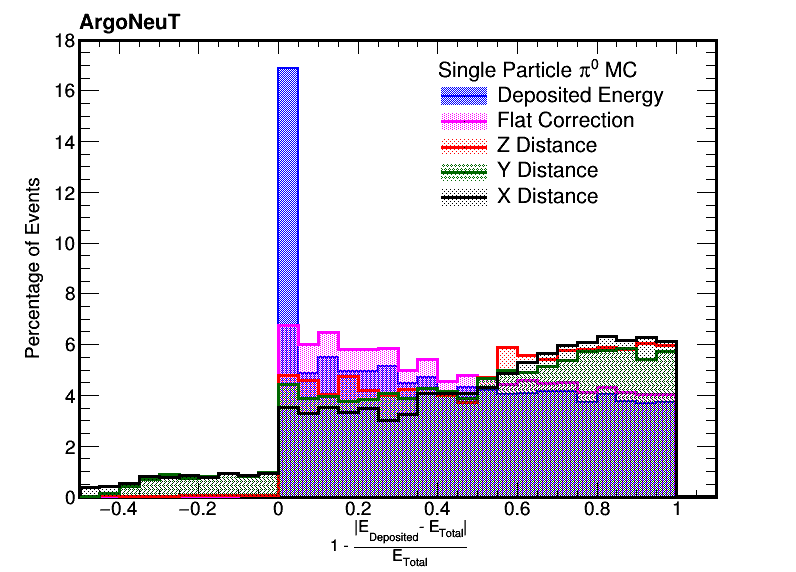}
    \caption{The fraction of energy from simulated $\pi^{0} \rightarrow \gamma\gamma$ events that after the application of the template based energy corrections. (Top) The blue distribution is the initial fraction of deposited energy inside the TPC and black is the result after the application of all the corrections. (Bottom) Shows the same fraction of deposited energy as the flat energy correction (pink), Z distance (red), Y distance (green), and X distance (black) is applied. } \label{fig:EnergyDepositedCorr}
 \end{center}
\end{figure}

The results of the application of the template based energy corrections and procedure described above shown in Fig. \ref{fig:EnergyDepositedCorr}. With each subsequent correction, more events move towards full containment (closer to 1) while only a relatively small fraction of the photons  ever have their energy corrected above the MC-truth value (these photons appear with values less than 0). The order of the corrections applied here start with the ``flat correction'' shown in pink (described in Section \ref{sec:LinearCorrAppend}), then the ``Z distance'' shown in red, ``Y distance'' shown in green, and ``X distance'' shown in black.

In order to ensure none of these templates would sculpt a result based on the $\pi^{0} \rightarrow \gamma\gamma$ system, the behavior of the templates themselves were studied for a sample of single particle photons and single particle electrons. These events were simulated with the same initial momentum and position distribution as the $\pi^{0}$ system and the topological templates were derived for the electron sample. Comparisons to  the fits from the electron and photon templates against those derived for the $\pi^{0} \rightarrow \gamma\gamma$ sample showed very little difference in either their shape of the magnitude of the correction. This is as expected since the development of the shower inside the detector has little to do with how the shower was created and more to do with the physical process of the electromagnetic properties of the argon.

In addition to checking the templates against single electrons and photon MC, a check that the order in which the templates are applied/derived was also performed. In the analysis the corrections are applied in the order described in Section \ref{sec:EnergyCorr}. The results were also computed by changing the order of the topological templates (from Z, Y, X distances to the boundary to all possible unique rearrangements of these three variables). The flat momentum based correction was always applied first, however the resulting templates were shown to be insensitive (up to a scaling) to which order they are applied.

\begin{center}
\begin{figure}[htb]
  \begin{center}
    \includegraphics[width=0.40\textwidth]{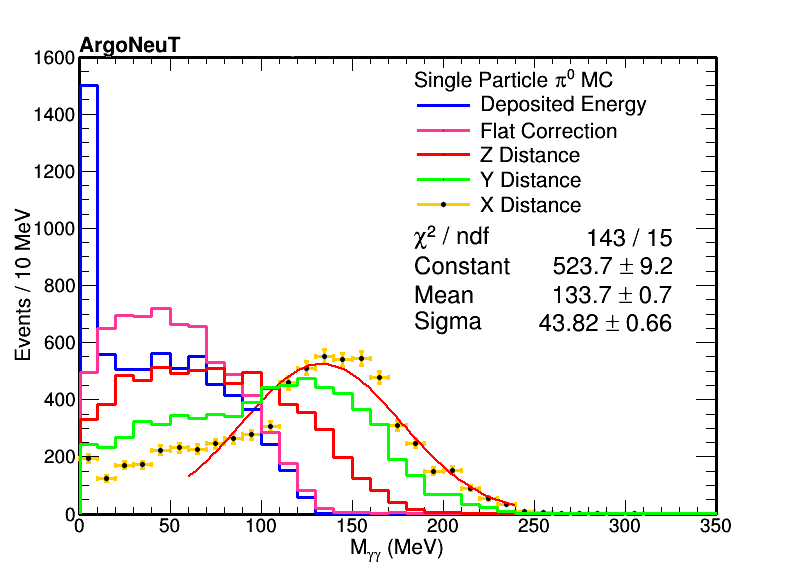}
    \includegraphics[width=0.40\textwidth]{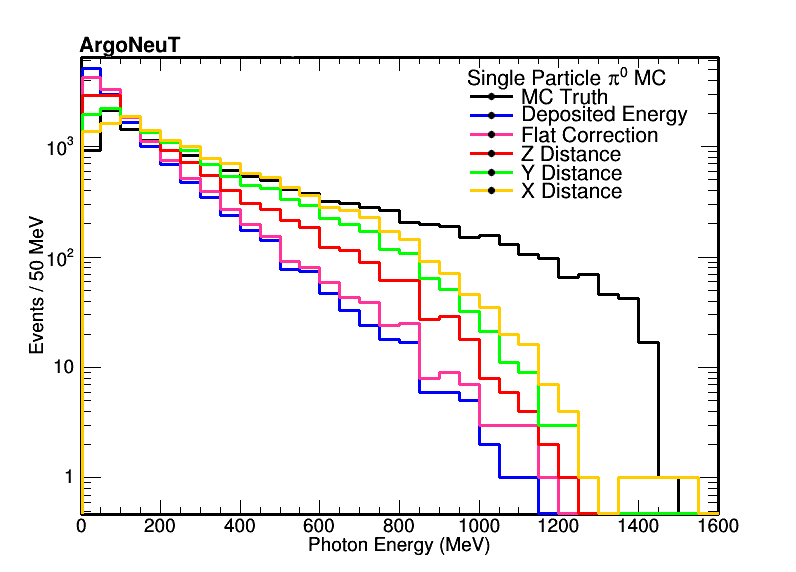}
    \includegraphics[width=0.40\textwidth]{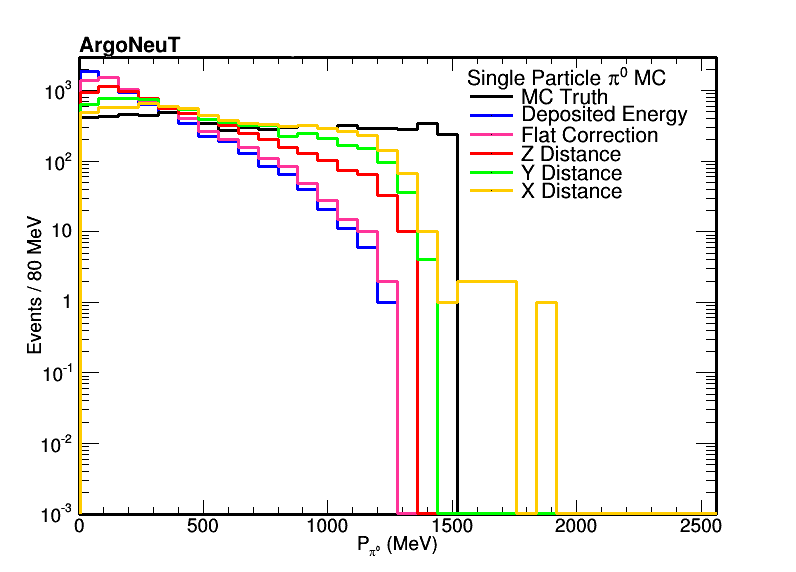}
    \caption{The invariant mass of the $\pi^{0} \rightarrow \gamma\gamma$ events fit with a Gaussian function (after applying all energy corrections) between 60~MeV$\leq M_{\gamma\gamma}\leq$240~MeV, the energy of the photons as well as $P_{\pi^{0}}$, calculated using Eq. \ref{eqn:EnergyMomentum}, for the  $\pi^{0} \rightarrow \gamma\gamma$ events that after the application of the template based energy corrections. } \label{fig:EnergyMomentumCorr}
 \end{center}
\end{figure}
\end{center}

Fig. \ref{fig:EnergyMomentumCorr} shows the impact of the template based energy correction procedure to the invariant mass of the $\pi^{0}$, the energy of the photons, as well as the momentum of the $\pi^{0}$. In black is shown the true distribution, in blue the information from only the deposited energy within the volume of the TPC, and then the application of each of the subsequent energy corrections.

\end{document}